\documentclass[12pt,a4paper]{article}

\usepackage[T1]{fontenc}
\usepackage[latin1]{inputenc}
\usepackage[nosort]{cite}
\usepackage{color}
\usepackage[bulletsep]{collref}
\usepackage{graphicx}
\usepackage{bbm}
\usepackage{amsmath}
\usepackage{amssymb}
\usepackage{physics}
\usepackage{mathtools}
\usepackage{subfig}
\usepackage{verbatim}
\usepackage{multirow}
\usepackage{tikz}
\usepackage{amsthm}
\usepackage{fancyhdr}
\usepackage{wrapfig}
\usepackage{hyperref}
\usepackage{dsfont}
\usepackage{delimset} 
\usepackage{shuffle} 
\usepackage{mathtools}

\usepackage{csquotes}
\usepackage{bm}
\usepackage{mathrsfs}  

\makeatletter \@addtoreset{equation}{section} \makeatother
\renewcommand{\theequation}{\thesection.\arabic{equation}}

\makeatletter
\let\old@startsection=\@startsection
\let\oldl@section=\l@section
\renewcommand{\@startsection}[6]{\old@startsection{#1}{#2}{#3}{#4}{#5}{#6\mathversion{bold}}}
\renewcommand{\l@section}[2]{\oldl@section{\mathversion{bold}#1}{#2}}
\makeatother

\makeatletter
\let\old@makecaption=\@makecaption
\def\@makecaption{\small\old@makecaption}
\makeatother

\renewcommand{\theequation}{\thesection.\arabic{equation}}

\let\oldPhi=\Phi
\let\oldPsi=\Psi
\let\oldGamma=\Gamma
\let\oldDelta=\Delta
\let\oldSigma=\Sigma
\let\oldTheta=\Theta
\let\oldPi=\Pi
\let\oldUpsilon=\Upsilon
\renewcommand{\Phi}{\mathnormal{\oldPhi}}
\renewcommand{\Psi}{\mathnormal{\oldPsi}}
\renewcommand{\Gamma}{\mathnormal{\oldGamma}}
\renewcommand{\Sigma}{\mathnormal{\oldSigma}}
\renewcommand{\Delta}{\mathnormal{\oldDelta}}
\renewcommand{\Theta}{\mathnormal{\oldTheta}}
\renewcommand{\Pi}{\mathnormal{\oldPi}}
\renewcommand{\Upsilon}{\mathnormal{\oldUpsilon}}

\newcommand{\Li}{\mathrm{Li}}

\newcommand{\zb}{\Bar{z}}
\newcommand{\thetab}{\Bar{\theta}}

\newcommand{\dimOp}{\mathrm{R}_D}
\newcommand{\loopOp}{\mathrm{R}_L}

\makeatletter
\newlength{\apb@width}
\newcommand{\autoparbox}[2][c]{\settowidth{\apb@width}{#2}\parbox[#1]{\apb@width}{#2}}
\newcommand{\includegraphicsbox}[2][]{\autoparbox{\includegraphics[#1]{#2}}}
\makeatother

\ifx\genfrac\sdflkaj

\else

\fi

\makeatletter
\def\mr@ignsp#1 {\ifx\:#1\@empty\else #1\expandafter\mr@ignsp\fi}%
\newcommand{\multiref}[1]{\begingroup
\xdef\mr@no@sparg{\expandafter\mr@ignsp#1 \: }%
\def\mr@comma{}%
\@for\mr@refs:=\mr@no@sparg\do{\mr@comma\def\mr@comma{,}\ref{\mr@refs}}%
\endgroup}
\makeatother

\newcommand{\hypref}[2]{\ifx\href\asklfhas #2\else\href{#1}{#2}\fi}
\newcommand{\Secref}[1]{Section~\multiref{#1}}

\newcommand{\appref}[1]{appendix~\multiref{#1}}
\newcommand{\Tabref}[1]{Table~\multiref{#1}}

\newcommand{\Figref}[1]{Figure~\multiref{#1}}
\renewcommand{\eqref}[1]{(\multiref{#1})}

\ifx\href\asklfhas\newcommand{\href}[2]{#2}\fi

\newcommand{\be}{\begin{eqnarray}}
\newcommand{\ee}{\end{eqnarray}}

\newcommand*\pFq[5]{{}_{#1}F_{#2}\left[\genfrac{}{}{0pt}{1}{#3}{#4};#5\right]}
\newcommand*\pcFq[5]{{}_{#1}\mathcal{F}_{#2}\left[\genfrac{}{}{0pt}{1}{#3}{#4};#5\right]}
\newcommand*\psvFq[5]{{}_{#1}\mathcal{F}^{\mathrm{sv}}_{#2}\left[\genfrac{}{}{0pt}{1}{#3}{#4};#5\right]}

\newcommand*\FF[4]{F_{#1}\left[\genfrac{}{}{0pt}{1}{#2}{#3};#4\right]}

\newcommand{\fs}{\mathfrak{s}}

\begin{document}

\thispagestyle{empty}

\begin{flushright}\footnotesize
\texttt{BONN-TH-2024-13}
\end{flushright}
\vspace{.2cm}

\begin{center}%
{\LARGE\textbf{\mathversion{bold}%
Conformal Four-Point Integrals: Recursive Structure, Toda Equations and Double Copy}\par}

\vspace{1cm}
{\textsc{Florian Loebbert, Sven F. Stawinski }}
\vspace{8mm} \\
\textit{%
Bethe Center for Theoretical Physics \\
Universit\"at Bonn, 53115, Germany
}
\vspace{.5cm}

\texttt{\{loebbert,sstawins\}@uni-bonn.de}
\vspace{1cm}
%

\par\vspace{15mm}

\textbf{Abstract} \vspace{5mm}

\begin{minipage}{12.7cm}
We consider conformal four-point Feynman integrals to investigate how much of their mathematical structure in two spacetime dimensions carries over to higher dimensions. In particular, we discuss recursions in the loop order and spacetime dimension. This results e.g.\ in new expressions for conformal ladder integrals with generic propagator powers in all even dimensions and allows us to lift results on 2d Feynman integrals with underlying Calabi--Yau geometry to higher dimensions. Moreover, we demonstrate that the Basso--Dixon generalizations of these integrals obey different variants of the Toda equations of motion, thus establishing a connection to classical integrability and the family of so-called tau-functions. We then show that all of these integrals can be written in a double copy form that combines holomorphic and anti-holomorphic building blocks. Here integrals in higher dimensions are constructed from an intersection pairing of two-dimensional ``periods'' together with their derivatives. Finally, we comment on extensions to higher-point integrals which provide a richer kinematical setup.
\end{minipage}
\end{center}

\newpage 

\tableofcontents
\bigskip
\hrule

\section{Introduction and Overview}
\label{sec:intro}

What is the difference between physics in two and four spacetime dimensions? On the one hand, it is well known that two dimensions are distinguished by their rich mathematical structure and computational accessibility. On the other hand, approaching particle phenomenology typically requires the calculation of observables in higher dimensions. It is thus a natural question in how far the investigation of two-dimensional objects contributes towards understanding four-dimensional physics.
In this paper we focus on a specific corner of this puzzle and ask how much of the structure of conformal four-point integrals in higher dimensions is contained in their two-dimensional counterparts.
\medskip

The interrelation between different mathematical structures becomes particularly apparent in the context of so-called fishnet integrals \cite{Zamolodchikov:1980mb}. These are representatives of Feynman integrals and at the same time correspond to full correlation functions within the fishnet conformal field theories introduced in \cite{Gurdogan:2015csr,Caetano:2016ydc,Kazakov:2018qbr,Kazakov:2018gcy,Loebbert:2020tje,Kazakov:2022dbd,Alfimov:2023vev}.  As such, they are ideally suited to transfer insights and methods among different research areas, in particular between the Feynman integral technology developed for the prediction of collider experiments and the concepts employed within conformal field theories. The most prominent examples of fishnet integrals correspond to Feynman graphs that can be cut out of a square lattice representing scalar interactions with four-point vertices.
\medskip

\begin{figure}[t]
    \centering
    \includegraphicsbox[width=.19\textwidth]{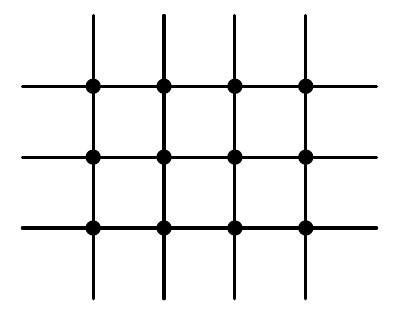}
     \includegraphicsbox[width=.19\textwidth]{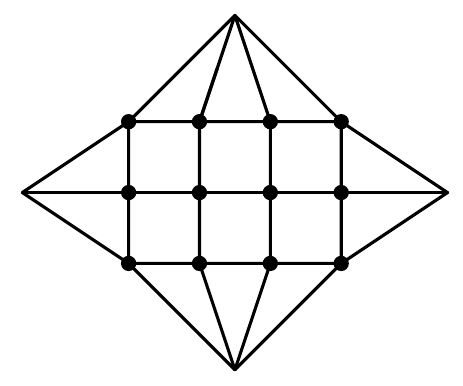}
    \includegraphicsbox[width=.19\textwidth]{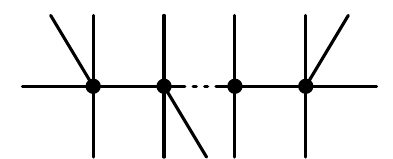}
    \includegraphicsbox[width=.19\textwidth]{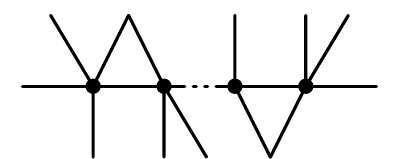}
    \includegraphicsbox[width=.19\textwidth]{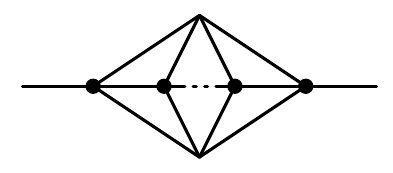}
    \caption{Examples of different types of (position space) Feynman diagrams discussed in this paper: a)~Fishnet, b)~four-point fishnet alias Basso--Dixon integral, c)~track-like diagram, d)~track with two coincidence limits, e)~ladder integral.}
    \label{fig:DiagramsIntro}
\end{figure}

For the special situation of four external points, fishnet integrals include the family of ladder integrals and their Basso--Dixon generalizations, which have been investigated from different directions, cf.\ \Figref{fig:DiagramsIntro}.
The classic result of Davydychev and Usyukina shows that ladder integrals in four spacetime dimensions with unit propagator powers can all be expressed in terms of single-valued polylogarithms generalizing the one-loop expression represented by the Bloch--Wigner dilogarithm \cite{Usyukina:1992jd,Usyukina:1993ch}. Recently, certain conformal ladder integrals in two spacetime dimensions were shown to be expressible as derivatives of ${}_pF_q$ hypergeometric functions \cite{Derkachov:2018rot}, which was generalized to 2d ladders with generic propagator powers in \cite{Duhr:2023bku}. Extending the familiy of ladder integrals by further rows of integration loops (or vertices in dual position space), leads to the so-called Basso--Dixon integrals \cite{Basso:2017jwq}. These have been found to evaluate to determinants of polylogs in four dimensions \cite{Basso:2017jwq} and of more general ladder integrals in two dimensions \cite{Derkachov:2018rot}.
\medskip

One of the reasons why fishnet diagrams are particularly suited for making progress on understanding multi-loop integrals is given by their integrable structures. As demonstrated by Zamolodchikov already in 1980, they provide a natural framework for defining statistical vertex models with an inherent notion of the quantum Yang-Baxter equation \cite{Zamolodchikov:1980mb}. More recently, a connection to Anti-de-Sitter/Conformal Field Theory integrability was drawn via the fishnet limit of gamma-deformed $\mathcal{N}=4$ super Yang-Mills theory \cite{Gurdogan:2015csr}. In particular, integrability allows for the application of new computational approaches to Feynman integrals, such as the separation of variables (SOV) method \cite{Sklyanin:1995bm,Derkachov:2018rot,Aprile:2023gnh,Alfimov:2023vev} or a bootstrap approach via their conformal Yangian symmetry \cite{Loebbert:2019vcj,Loebbert:2020glj,Corcoran:2020epz,Duhr:2022pch}. It was also instrumental in identifying the above determinant structures of Basso--Dixon type \cite{Basso:2017jwq}.
\medskip

A setup which is mathematically particularly rich is given by the class of isotropic fishnet integrals in two spacetime dimensions, with propagator powers adapted to certain rational values that guarantee conformal symmetry \cite{Duhr:2022pch,Duhr:2023eld,Duhr:2024hjf}. These integrals feature an underlying Calabi--Yau geometry that in turn is completely determined by their integrable structures in the form of Yangian symmetry.  While the connection to geometry and integrability implies remarkable control over this family of multi-loop Feynman integrals in 2d, from the physics perspective the situation clearly represents a toy model. With regard to the above discussion it is thus very natural to ask how much of this  mathematical control can be lifted to multi-loop integrals in higher spacetime dimensions which look more natural from the perspective of particle or gravitational wave physics.
\medskip

In the present paper we approach this question by focussing on conformal four-point integrals that are characterized by a function $\phi(z,\bar z)$ of two conformal variables $z$ and $\zb$. These integrals are distinguished by the fact that the number of conformal variables is independent of the spacetime dimension for $D>1$, which is no longer the case for higher numbers of external points. We note, however, that several of our findings also hold beyond four points with a restriction to limited kinematics. An important role within the present paper is played by two first order differential operators, which essentially generate rotations and dilatations on the two-dimensional plane defined by the effective coordinates $z$ and $\bar z$ after fixing conformal symmetry:
\begin{equation}
\dimOp=\frac{z\partial_z-\zb\partial_{\zb}}{z-\zb},
\qquad\quad
\loopOp=-\frac{z\partial_z+\zb\partial_{\zb}}{\log(z\zb)} .
\label{eq:dimAndLoopOps}
\end{equation}
In particular, we make the following observations:
\begin{itemize}
\item The class of \emph{track-like} Feynman integrals%
\footnote{These track-like integrals are generalizations of the so-called train track integrals referring to such Feynman graphs made out of four-point vertices only \cite{Bourjaily:2018ycu}. The name train track originates from the form of the graphs in dual momentum space.}  obeys a dimension shift relation in the form
\begin{equation}
\dimOp \,\phi^{(D)}=\phi^{(D+2)}.
\end{equation}
Here $\dimOp$ represents the simple first order differential operator given above and as opposed to standard dimension-shift relations for Feynman integrals, $\phi$ corresponds to a \emph{single} conformal integral on both sides of the equation. We define the family of track-like Feynman integrals as (position space) graphs composed of a chain of integration vertices with arbitary numbers of external points and glued to each other with each vertex being connected to at most two internal propagators. We also allow some of the external points to be identified as long as there is at least one external point connected to (and only to) each end of the chain of integration vertices, cf.\ \Figref{fig:DiagramsIntro}.
Note that only for four points the recursion can reach the full kinematic space of the higher dimensional integral on the right hand side, which makes the ladder integrals the prime example considered below. 
We note that the recursion holds for Feynman graphs with massless internal propagators, and massive or massless external propagators, with a suitable massive generalization of conformal variables and the differential operator.
\item As an application of the above dimensional recursion, we obtain new expressions for conformal ladder integrals with generic propagator powers in all even dimensions, and we lift 2d results on integrals associated to Calabi--Yau geometries to higher dimensions.
\item
The above first order differential operator $\loopOp$ induces a recursion on Basso--Dixon integrals (including the ladders) in four dimensions, which takes the form of a variant of the Toda equation of motion. Similarly, 2d four-point fishnet integrals obey a Toda equation. As such, Basso--Dixon integrals relate to tau-functions, which have been studied in a broader context as solutions of the bilinear equations of Hirota type, hinting at a connection to classical integrability.
\item Finally we emphasize that the considered four-point integrals in all even dimensions can be written in a double copy representation of the form
\begin{equation}
\phi=\Pi^\dagger(\zb)\Sigma\Pi(z),
\end{equation}
with a constant matrix $\Sigma$ and the vector $\Pi(z)$ consisting of 2d periods and their derivatives with respect to the holomorphic conformal variable(s) $z$.  Going beyond four external points, this factorization argument applies on a subslice of kinematic space.
\end{itemize}
Our findings generalize and combine a couple of recent insights concerning the above family of integrals, which are discussed in more detail in the subsequent sections and briefly summarized in the following.
\medskip

In particular, the operator $\dimOp$ is well known to relate scalar four-point conformal blocks in two and four spacetime dimensions \cite{Simmons-Duffin:2012juh}, as well as contact Witten diagrams with different external scaling dimensions \cite{DHoker:1999kzh,Rigatos:2022eos}.
In the present paper we emphasize its usefulness in the context of Feynman integral computations and establish a dimensional recursion for the class of track-like Feynman integrals (\Secref{subsec:confSchwingerParam}). We apply this recursion e.g.\ to obtain results for conformal ladder integrals with generic propagator powers in all even dimensions (\Secref{subsec:laddersInAllD}), or to lift results on the above mentioned 2d fishnet integrals with an underlying Calabi--Yau geometry to four dimensions (\Secref{subsec:CYlift}).
\medskip

Relatedly, the class of four-point ladder integrals for certain propagator powers was recently shown to map to twisted partition functions of massive free scalar fields \cite{Petkou:2021zhg,Karydas:2023ufs}. Here the differential operators $\dimOp$ and $\loopOp$ were employed to compare recursion relations among partition functions and ladder integrals in different dimensions or at different loop orders, respectively. Similar relations were found in the context of the SOV representation of those ladder integrals, which were shown to have a very similar form in two and four dimensions \cite{Derkachov:2018rot,Derkachov:2019tzo,Derkachov:2021ufp,Basso:2021omx}. In particular, in the present paper we show that the operator $\loopOp$ also has a natural action on Basso--Dixon integrals resulting in variants of bilinear Toda or Hirota equations as discussed in \Secref{sec:loopRecursion}, and that the action of $\dimOp$ on ladder integrals generalizes to arbitrary propagator powers.
\medskip

Concerning the appearence of Toda-like bilinear equations, we note that recently several observables in the context of supersymmetric gauge theories were found to feature a determinant form, see \cite{Beccaria:2022ypy,Bajnok:2024epf} and references therein. For instance, the so-called octagon form factor was shown to admit a determinant representation, which solves the 1d or 2d Toda equation in certain lightcone double-scaling limits \cite{Belitsky:2020qir,Olivucci:2021pss}. 
\medskip

Finally, it was recently shown that all Feynman integrals with non-integer propagator powers in two spacetime dimensions can be written in terms of single-valued analogues of Aomoto-Gelfand hypergeometric functions~\cite{Duhr:2023bku}. For track-like graphs we can employ the dimensional recursion given above, to lift these results  to four and higher even spacetime dimensions. For four external points this reproduces the full result for the higher dimensional integral while for higher number of points the result is constrained to a kinematic subspace. In particular, this implies a double copy form of the higher dimensional integrals as illustrated in \Secref{sec:doubleCopy}.
\medskip

The results discussed in the present paper lead to a plethora of research avenues relating the topics of conformal field theory, integrability and Feynman integrals on which we give a brief outlook in \Secref{sec:Outlook}.

\section{Dimensional Recursion for Conformal Integrals}
\label{sec:dimRecursion}

In this section we will establish a dimensional recursion for conformal track-like integrals. This will for instance allow us to relate a generic conformal ladder in any even dimension to the known 2d result by the (repeated) action of a first order differential operator. Fixing the propagator powers to specific values, this statement in particular applies to the class of 2d ladder integrals related to Calabi--Yau varieties that can be lifted to four (or higher) dimensions in this way. We will further explain how to take the somewhat subtle polylogarithmic limit of unit propagator powers and point out some connections to known dimensional recursions of conformal blocks and Witten diagrams.
\subsection{Conformal Four-Point Feynman Integrals}
\label{subsec:fourierMellin}
The main object of study in this paper are conformal four-point Feynman integrals of the form
\begin{equation}
    I_{\mathbf{a}}(x_1,\dots,x_4)=\int\prod_{i=1}^L\frac{\mathrm{d}^D y_i}{\pi^{D/2}}\prod_{(j,k)\in T_1}\frac{1}{(y_j-y_k)^{2a_{j,k}}}\prod_{(l,m)\in T_2}\frac{1}{(x_l-y_m)^{2a_{l,m}}}.
\end{equation}
The integrals depend on four external points $x_1,\dots x_4$ and a set of propagator weights $\mathbf{a}$ as well as (implicitly) on the spacetime dimension $D$, the loop order $L$ and 
the index sets $T_1,T_2$ determining the topology of the corresponding Feynman graph. The conformal symmetry is ensured by requiring that the propagator weights corresponding to internal and external edges attached to a fixed integration vertex sum up to the spacetime dimension $D$:
\begin{equation}
    \sum_{j:\,(j,k)\in T_1}a_{j,k}+\sum_{l:\,(l,k)\in T_2}a_{l,k}=D,\quad \text{for all }k.
\end{equation}
The integrals we are studying here can have at least two different interpretations:
\begin{itemize}
    \item They may arise in the computation of correlation functions, in which case the external points represent spacetime coordinates and the conformal symmetry is the usual spacetime conformal symmetry. 
    \item They can be interpreted as integrals contributing to some (off-shell) planar scattering amplitude in which case the external points represent so-called dual or region momenta, related to the momenta in the scattering process through $p_i=x_{i}-x_{i+1}$, and the conformal symmetry corresponds to what is usually called dual conformal symmetry \cite{Drummond:2006rz,Drummond:2008vq}. 
\end{itemize}
    In the following we will just speak of conformal symmetry and not further differentiate between the two possible interpretations given above. Note however, that we will denote the number of integration points as the number of loops, in accord with the scattering amplitudes picture. 

\paragraph{Conformal Symmetry for Four Points.}

\begin{table}[t]
\begin{center}
\begin{tabular}{|l||c|c|c|c|c|c|c|c|}
\hline
     Number of points $n$ & 2&3&4&5&6&7&8&\dots  
     \\\hline
     Number of cross ratios in $D=2$& 0&0&2&4&6&8&10&\dots
     \\
    Number of cross ratios in generic $D$& 0&0&2&5&9&14&20&\dots
      \\
      \hline
\end{tabular}
\end{center}
\caption{The four-point configuration is distinguished by the fact that $2$-dimensional kinematics agrees with the kinematics in general dimensions $D$.}
\label{tab:crossratiosnpoints}
\end{table}

As is well known we can make use of conformal symmetry to fix three of the external points $x_i$ and rotate the last point of a four-point configuration into a two-dimensional plane. The integral can hence only depend on a point in this plane that is parametrized by two numbers $z,\zb$ which can be related to the original coordinates via the conformal cross ratios
\begin{equation}
\label{eq:conformalCrossRatios}
    u=\frac{x_{12}^2x_{34}^2}{x_{14}^2x_{23}^2}=z\zb,
    \qquad v=\frac{x_{13}^2x_{24}^2}{x_{14}^2x_{23}^2}=(1-z)(1-\zb).
\end{equation}
Depending on the kinematics, $z$ and $\bar z$  are complex conjugate to each other or real and independent. 

In practice this means that we can split off a prefactor $V_\mathbf{a}(x_1,\dots,x_4)$ carrying the conformal weights, leaving us with a function only depending on the two variables $z,\zb$:
\begin{equation}
    I_{\mathbf{a}}(x_1,\dots x_4)=V_{\mathbf{a}}(x_1,\dots x_4)\phi_{\mathbf{a}}(z,\zb).
\end{equation}
Importantly for the present paper, these remaining degrees of freedom $z,\zb$ are precisely the same as those of a four-point correlator in two-dimensional CFT, i.e.\ the numbers of independent cross ratios in two and generic dimensions agree: $2(n-3)=n(n-3)/2$ for $n=4$ external points. Note that this is no longer true for $n>4$, cf.\ \Tabref{tab:crossratiosnpoints}.

\paragraph{Ladder Integrals.}

\begin{figure}[t]
\begin{center}
\includegraphicsbox[]{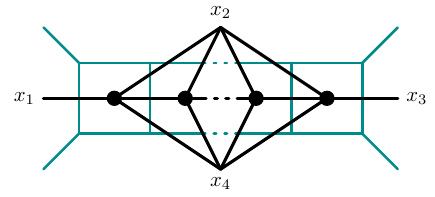}
\end{center}
\caption{Four-point ladder integrals in position space with coordinates $x_j^\mu$ (black) and dual momentum space with momenta $p_j^\mu=x_j^\mu-x_{j+1}^\mu$ (green).}
\label{fig:ladders}
\end{figure}

One particular class of conformal four-point integrals that we will repeatedly study in this paper are the conformal ladder integrals defined as
\begin{align}
    I_{D;\mathbf{a}}^{(L)}&=\int\prod_{i=1}^L\frac{\mathrm{d}^Dy_i}{\pi^{D/2}}\,\frac{1}{(x_1-y_1)^{2 a_1}}\frac{1}{(x_3-y_L)^{2a_{L+1}}}\prod_{i=1}^{L-1}\frac{1}{(y_i-y_{i+1})^{2 a_{i+1}}} \nonumber\\
    &\qquad\qquad\times\prod_{i=1}^L\frac{1}{(x_2-y_i)^{2a_{L+1+i}}}\prod_{i=1}^L\frac{1}{(x_4-y_i)^{2a_{2L+1+i}}}, 
    \label{eq:defLadderIntegrals}
\end{align}
see \Figref{fig:ladders}, with the propagator weights satisfying the conformal constraints
\begin{equation}
    a_i+a_{i+1}+a_{L+i+1}+a_{2L+i+1}=D,\qquad i=1,\dots,L \,.
\end{equation}
As remarked above, conformal symmetry lets us split off a factor
\begin{equation}
    I_{D;\mathbf{a}}^{(L)}=V_{D;\mathbf{a}}^{(L)} \,\phi_{D;\mathbf{a}}^{(L)}(z,\zb),
\end{equation}
which we choose as
\begin{align}
    \label{eq:VFactorLadder}
    V_{D;\mathbf{a}}^{(L)}&=\left(x_{14}^2\right)^{-a_1}\left(x_{23}^2\right)^{(-a_1-a_{L+1}-(a_{L+2}+\dots+a_{2L+1})+(a_{2L+2}+\dots +a_{3L+1}))/2} \\
    &\quad\times  \left(x_{24}^2\right)^{(a_1+a_{L+1}-(a_{L+2}+\dots+a_{2L+1})-(a_{2L+2}+\dots +a_{3L+1}))/2} \nonumber \\
    &\quad \times \left(x_{34}^2\right)^{(a_1-a_{L+1}+(a_{L+2}+\dots+a_{2L+1})-(a_{2L+2}+\dots +a_{3L+1}))/2}\left(\frac{x_{12}^2x_{34}^2}{x_{14}^2x_{23}^2} \right)^{LD/2-\sum_{i=1}^{2L+1}a_i+a_{L+1}}.
\end{align}
This choice of prefactor $V_{D;\mathbf{a}}^{(L)}$ and the variables $z,\zb$ in \eqref{eq:conformalCrossRatios} is convenient to obtain a compact formula for ladder integrals in any even dimension.

\subsection{Conformal Schwinger Parametrization and Dimensional Recursion}
\label{subsec:confSchwingerParam}

In this subsection we will derive a simple dimensional recursion relation for track-like conformal Feynman integrals. The existence of this recursion will be made clear through a conformal version of the familiar Schwinger parametrisation \cite{Symanzik:1972wj}. This representation has recently been employed in the study of one- and two-loop conformal Feynman integrals \cite{Paulos:2012nu} as well as Witten diagrams \cite{Rastelli:2017ecj}. We will review the derivation here, following \cite{Paulos:2012nu}, and then extend it to all-loop track-like graphs.%

The following derivation is best done in embedding space \cite{Dirac:1936fq,Weinberg:2010fx,Simmons-Duffin:2012juh}. We will hence represent the external and internal points $x_i,y_i$ in Euclidean space $\mathbb{R}^D$ by points $X_i, Y_i$ on the null cone of the real projective space $\mathbb{PR}^{D+1}$ with Minkowski signature $(+,\dots,+,-)$ such that propagators linearize
\begin{equation}
    (x_i-y_j)^2=-X_i\cdot Y_j,
\end{equation}
for more details we refer to \cite{Simmons-Duffin:2012juh}. We will now demonstrate the key steps in the derivation for a general one-loop integral and then generalize to any track-like diagram obtained by gluing any number of one-loop diagrams in a chain-like (or track-like) manner.

\paragraph{One-loop Diagrams.}

A general one-loop integral has the form\footnote{For details on the integration measure we refer to \cite{Simmons-Duffin:2012juh}.}
\begin{equation}
    I^{(1)}_\mathrm{gen}=\int\dd Y \prod_{i=1}^n\frac{1}{(-X_i\cdot Y)^{a_i}},
\end{equation}
where the propagator powers satisfy the conformal constraint
\begin{equation}
    \sum_{k=1}^n a_k=D.
    \label{eq:confconst1}
\end{equation}
Schwinger parametrising and solving the Gaussian $Y$-integration yields
\begin{equation}
    I^{(1)}_\mathrm{gen}=\frac{1}{\prod_{i=1}^n\Gamma(a_i)}\prod_{i=1}^n\left( \int_0^\infty\dd t_i\, t_i^{a_i-1}\right)\left( \sum_{j=1}^n t_j \right)^{-h}\exp\left[ \frac{(\sum_{k=1}^n t_k X_k)^2}{2\sum_{l=1}^n t_l} \right],
\end{equation}
with $h=D/2$. Introducing an auxiliary 1 in the form 
\begin{equation}
    1=\int_{0}^\infty \delta\left(v-\sum_{i=1}^n t_i \right),
\end{equation}
and rescaling $t_i\rightarrow v t_i$ we find
\begin{equation}
    I^{(1)}_\mathrm{gen}=\frac{1}{\prod_{i=1}^n\Gamma(a_i)}\prod_{i=1}^n\left( \int_0^\infty\dd t_i\, t_i^{a_i-1}\right)\int_0^\infty\dd v \,\delta\left(1-\sum_{i=1}^n t_i \right) v^{\sum_{i=1}^n a_i-h-1} e^{\frac{v}{2}\left(\sum_{i=1}^n t_i X_i\right)^2 }\,.
\end{equation}
Further rescaling $t_i\rightarrow v^{-1/2} t_i$, the integral simplifies due to the conformal constraint \eqref{eq:confconst1} and we find 
\begin{equation}
    I^{(1)}_\mathrm{gen}=\frac{2}{\prod_{i=1}^n\Gamma(a_i)}\left( \int_0^\infty\dd t_i\, t_i^{a_i-1}\right)\exp\left[ \frac{1}{2}\left(\sum_{i=1}^n t_i X_i\right)^2 \right],
\end{equation}
or written in terms of $D$-dimensional quantities
\begin{equation}
    I^{(1)}_\mathrm{gen}=\frac{2}{\prod_{i=1}^n\Gamma(a_i)}\left( \int_0^\infty\dd t_i\, t_i^{a_i-1}\right)\exp\left[ -\sum_{i<j}t_i t_j x_{ij}^2 \right].
\end{equation}
Clearly a derivative with respect to some squared distance $x_{ij}^2$ leads to a shift of two propagator powers by one and hence, due to the conformal constraint, to a shift in the dimension by two.  

\paragraph{Track-like Diagrams.}

In \cite{Paulos:2012nu} the authors generalized this representation to  two-loop integrals, which we will now further generalize  to arbitrary track-like integrals. 
A general $L$-loop track-like integral takes the form
\begin{align}
    I_\mathrm{gen}^{(L)}&=\int\dd Y_1\dots\int\dd Y_L\left(\prod_{i\in S_1} \frac{1}{(-X_i\cdot Y_1)^{a_i^{(1)}}} \right)\frac{1}{(-Y_1\cdot Y_2)^{b_1}}\left(\prod_{i\in S_2} \frac{1}{(-X_i\cdot Y_2)^{a_i^{(2)}}} \right) \nonumber\\
    &\hspace{5cm}\dots\frac{1}{(-Y_{L-1}\cdot Y_L)^{b_{L-1}}}\left(\prod_{i\in S_L} \frac{1}{(-X_i\cdot Y_L)^{a_i^{(L)}}} \right),
    \label{eq:test}
\end{align}
with $b_0=b_L=0$ and the conformal conditions
\begin{equation}
    b_{i-1}+b_i+\sum_{j\in S_i}a_j^{(i)}=D,\qquad i=1,\dots,L.
\end{equation}
Here $S_1,\dots,S_L$ are some index sets keeping track of which external points are connected to which integration point. Note that our definition of track-like integrals as given in \Secref{sec:intro} demands that there exist two external points that are only in $S_1$ and $S_L$, respectively.

Schwinger parametrising the integral yields
\begin{align}
    I_\mathrm{gen}^{(L)}&=\int\dd Y_1\dots\int\dd Y_L\int[\dd t]\int[\dd s]\exp(P_1\cdot Y_1+\dots P_L\cdot Y_L) \nonumber\\
    &\qquad\qquad\times\exp(s_1 Y_1\cdot Y_2+s_2 Y_2\cdot Y_3+\dots +s_{L-1}Y_{L-1}\cdot Y_L) ,
\end{align}
with the abbreviations
\begin{align}
    P_i&=\sum_{j\in S_i}t_{i,j} X_j,
    &
    [\dd t]&=\prod_{i=1}^L\prod_{j\in S_i}\frac{\dd t_{i,j}\,t_{i,j}^{a_j^{(i)}-1}}{\Gamma(a_j^{(i)})},
    &
    [\dd s]&=\prod_{i=1}^{L-1}\frac{\dd s_i\,s_i^{b_i-1}}{\Gamma(b_i)}.
\end{align}
In the exponent every integration point $Y_i$ mutliplies a linear combination of external and internal points with the coefficients given by some Schwinger parameters. The conformal condition then forces the propagator powers associated with these Schwinger parameters to sum up to the dimension $D$ for every internal point $Y_i$.

We claim that if this conformal condition holds, then we can evaluate all $Y$-integrals to find the conformal Schwinger parametrisation
\begin{align}
    I_\mathrm{gen}^{(L)}&=2^L\int[\dd t]\int[\dd s]\exp\left(\frac{1}{2}P_1^2\right)\exp\left(\frac{1}{2}(s_1 P_1+P_2)^2\right)\exp\left(\frac{1}{2}(s_1 s_2 P_1+s_2 P_2+P_3)^2\right) \nonumber\\
    &\qquad\qquad\times\dots\exp\left(\frac{1}{2}(s_1 \dots s_{L-1} P_1+s_2\dots s_{L-1} P_2+\dots+ s_{L-1}P_{L-1}+P_L)^2\right).
\end{align}
We will prove this by induction over the number of $Y$-integrals.%
\footnote{Note that this is not really an induction over the loop order $L$ since we keep $L$ in the $t$- and $s$-integration measures fixed.}
We already saw that the formula is correct for one $Y$-integration. To perform the inductive step we will follow the same steps as before to perform the first $Y$-integration and find
\begin{align*}
    I_\mathrm{gen}^{(L)}&=2\int\dd Y_2\dots\int\dd Y_L\int[\dd t]\int[\dd s]\exp\left(\frac{1}{2}(P_1+s_1 Y_2)^2\right)\exp(P_2\cdot Y_2+\dots P_L\cdot Y_L) \\
    &\qquad\qquad\times\dots\exp(s_2 Y_2\cdot Y_3+\dots +s_{L-1}Y_{L-1}\cdot Y_L) \\
    &=2\int\dd Y_2\dots\int\dd Y_L\int[\dd t]\int[\dd s]\exp\left(\frac{1}{2}P_1^2\right)\exp((s_1P_1+P_2)\cdot Y_2+\dots P_L\cdot Y_L) \\
    &\qquad\qquad\times\dots\exp(s_2 Y_2\cdot Y_3+\dots +s_{L-1}Y_{L-1}\cdot Y_L).
    \addtocounter{equation}{1}\tag{\theequation}
\end{align*}
Here we used that the $Y_i$ are null, in particular $Y_1^2=0$, which reflects the fact that the internal propagators are massless. The key point for the induction is now that this is indeed of the same form as above if we define
\begin{equation}
    P_2'=s_1 P_1+\sum_{j\in S_2}t_{2,j}X_j,
    \label{eq:P2prime}
\end{equation}
where we view $P_1$ as some fixed vector. In particular in evaluating further integrals we would not rescale the parameters $t_{1,j},\, j\in S_1$ anymore. Then as before the conformal condition is satisfied, since for example the Schwinger parameters in the coefficient of $Y_2$ are $s_1,s_2,t_{2,j},\, j\in S_2$, such that the associated sum over propagator weights is indeed equal to the spacetime dimension: 
\begin{equation}
    b_1+b_2+\sum_{j\in S_2}a_j^{(2)}=D.
\end{equation}
Applying the induction hypothesis now yields
\begin{align}
    I_\mathrm{gen}^{(L)}&=2^L\int[\dd t]\int[\dd s]\exp\left(\frac{1}{2}P_1^2\right)\exp\left(\frac{1}{2}P_2'^2\right)\exp\left(\frac{1}{2}(s_2 P_2'+P_3)^2 \right)
    \nonumber\\
    &\qquad\qquad\times \dots\exp\left( \frac{1}{2}(s_2\dots s_{L-1}P_2'+s_3\dots s_{L-1}P_3+\dots +s_{L-1}P_{L-1}+P_L)^2 \right) ,
\end{align}
    such that plugging in \eqref{eq:P2prime} results in
    \begin{align}
    I_\mathrm{gen}^{(L)}&=2^L\int[\dd t]\int[\dd s]\exp\left(\frac{1}{2}P_1^2\right)\exp\left(\frac{1}{2}(s_1 P_1+P_2)^2\right)\exp\left(\frac{1}{2}(s_1 s_2 P_1+s_2 P_2+P_3)^2\right) \nonumber\\
    &\qquad\qquad\times \dots\exp\left(\frac{1}{2}(s_1 \dots s_{L-1} P_1+s_2\dots s_{L-1} P_2+\dots+ s_{L-1}P_{L-1}+P_L)^2\right)\,,
    \label{eq:confSchwingerTracklike}
\end{align}
as claimed.
\paragraph{Generic Conformal Integrals.}

The integral representation \eqref{eq:confSchwingerTracklike}, looks very similar to the usual Schwinger representation of Feynman integrals but with the first Symanzik polynomial absent, see e.g.~\cite{Weinzierl:2022eaz}. This is a consequence of conformal symmetry and we will hence refer to this representation as \emph{conformal Schwinger parametrisation}.\footnote{In its current form the representation seems to depend on all squared distances, but in any concrete example (after stripping off a prefactor carrying the conformal weights) it is simple to find a representation only depending on the conformal cross ratios by rescaling the Schwinger parameters analogous to \cite{Bourjaily:2019jrk}.} In our derivation we focussed on track-like integrals, but this was entirely for simplicity allowing us to be quite explicit while including a large set of integrals. This representation exists for any conformal Feynman integral and in any concrete application, it is straightforward to derive this representation, following exactly the same steps as before.
\paragraph{Massive Extension.}

Furthermore we assumed that all propagators were massless which is reflected by the fact that all points in embedding space $X_i,Y_i$ lie on the null cone. However, in the derivation only the null condition on the internal points was used, such that the above derivation can be extended to graphs with external propagators being massive without changing any of the arguments. This is analogous to the findings of \cite{Loebbert:2020hxk}, where it was conjectured that the Yangian symmetry of fishnet Feynman integrals generalizes to a massive Yangian symmetry if only the boundary propagators are massive. This massive Yangian symmetry is the infinite-dimensional extension of the massive dual conformal symmetry already noticed in the 50s  \cite{wickMassiveConformal,cutkoskyMassiveConformal} and more recently studied in the context of the AdS/CFT correspondence \cite{Alday:2009zm,Caron-Huot:2014gia}.

\paragraph{Dimensional Recursion.}

From the conformal Schwinger parametrisation in equation \eqref{eq:confSchwingerTracklike}, we can now easily see the dimensional recursion. A derivative with respect to a kinematic invariant $x_{ij}^2$ pulls down factors of the integration variables, which effectively leads to a shift in the propagator powers and hence, by the conformal constraints, in the dimension.

A priori a kinematic invariant could multiply multiple terms leading to a sum of integrals with differently shifted dimensions, which might not even individually be conformally invariant. It could also happen that the kinematic invariant is multiplied by only one string of Schwinger parameters, which however, after the differentiation with respect to the kinematic invariant, lead to shifts in the propagator weights which do not preserve the conformal condition at all vertices.

For track-like integrals we however find a simple, manifestly conformal recursion mapping one conformal integral to another with the dimension increased by two. This can be seen from equation \eqref{eq:confSchwingerTracklike} as follows. By definition there exist two external points $x_i,x_j$ such that $i\in S_1,j\in S_L$ and they appear in no other index sets. Hence the kinematic invariant $x_{ij}^2$ only appears in the scalar product $P_1\cdot P_L$ which is multiplied by a single string of Schwinger parameters $s_1\dots s_{L-1}t_{1,i}t_{L,j}$. These correspond to the propagators which join the two external points $i,j$ along the chain of integration vertices. Hence every integration point is connected to precisely two propagators which are shifted and hence the conformal condition is preserved and simply leads to a dimension shift by two.

Note that also for more general conformal integrals the terms which multiply a certain kinematic invariant $x_{ij}^2$ in the conformal Schwinger parametrization seem to have an interpretation in terms of possible paths between the points $i$ and $j$. A better understanding of this could lead to simple graphical rules which allow to simply write down the conformal Schwinger parametrization for any conformal Feynman integral. The exploration of this however goes beyond the scope of this work.

For track-like integrals we can use the recursion to relate the integral in arbitrary dimension to a single integral in some lower dimension. This is particularly useful when we relate an even-dimensional integral to an integral in two dimensions, which is easier to compute since it takes a double-copy form, see  \Secref{sec:doubleCopy}. In particular, there are general results for some interesting classes of 2d integrals that can be lifted to higher dimensions in this way \cite{Duhr:2022pch,Duhr:2023bku,Duhr:2023eld,Duhr:2024hjf}. 

\paragraph{Beyond 4 Points.}
For $n>4$ points there is however a subtlety, namely that the kinematic space in lower dimensions is more restricted than in higher dimensions due to the existence of Gram constraints. Explicitly a conformal $n$-point integral depends on $n(n-3)/2$ cross-ratios in a sufficiently high dimension but only on $2(n-3)$ cross ratios in two dimensions. 

Let us make this fully explicit for five external points. In generic dimension we can form 5 cross-ratios out of the external points which can be chosen as (following \cite{Buric:2021ywo})
\begin{align}
    u_1&=\frac{x_{12}^2x_{34}^2}{x_{13}^2x_{24}^2}=z_1\zb_1\,, \qquad 
    v_1=\frac{x_{14}^2x_{23}^2}{x_{13}^2x_{24}^2}=(1-z_1)(1-\zb_1) \,,\nonumber\\
    u_2&=\frac{x_{23}^2x_{45}^2}{x_{24}^2x_{35}^2}=z_2\zb_2\,,\qquad 
    v_1=\frac{x_{25}^2x_{34}^2}{x_{24}^2x_{35}^2}=(1-z_2)(1-\zb_2)\,, \\
    U&=\frac{x_{15}^2x_{23}^2x_{34}^2}{x_{24}^2x_{13}^2x_{35}^2}=w(z_1-\zb_1)(z_2-\zb_2)+(1-z_1-z_2)(1-\zb_1-\zb_2)\,. \nonumber
\end{align}
The kinematic space here is given by two planes parametrized by $z_1,\zb_1$ and $z_2,\zb_2$ respectively, intersecting in some angle parametrized by $w$ \cite{Buric:2021kgy}. As explained in \cite{Buric:2021kgy}, if the five points lie in a plane, i.e.~we have two-dimensional kinematics, we have $w=0$ or $w=1$,%
\footnote{There is a $\mathbb{Z}_2\times\mathbb{Z}_2$ symmetry with generators acting like $z_r\leftrightarrow \zb_r,\, w\mapsto 1-w$ for $r=1,2$, hence $w=0$ and $w=1$ correspond to the same kinematic setup just constitute different parametrizations \cite{Buric:2021kgy}.} i.e.\ the two planes parametrized by $z_1,\zb_1$ and $z_2,\zb_2$ collapse into a single one. If we now have some conformal five-point integral in two dimensions, it is given by (after stripping off a factor carrying the conformal weights) a function $\phi(\bm{z},\bm{\zb})$ of the four variables $\bm{z}=(z_1,z_2),\bm{\zb}=(\zb_1,\zb_2)$. We can then find a dimensional recurrence from the conformal Schwinger parametrization which will be given by some first order differential operator in these four variables. The result of the differential recursion can hence clearly not depend on $w$ and only agrees with the four-dimensional result for $w=0$ or $w=1$, depending on the parametrization.

In summary, the conformal Schwinger parametrization exists for any conformal Feynman integral and will lead to a simple dimensional recursion if the integral is track-like. This allows us to relate the integral in arbitrary dimension to a lower-dimensional integral; relating it to a two-dimensional integral  without any kinematic restrictions, is however only possible for four-point integrals.

Let us illustrate the recursion on the familiar example of the four-point ladder integrals.


\subsection{Example: Parametric Ladders in All Even Dimensions}
\label{subsec:laddersInAllD}

In the previous subsection we have found a simple dimensional recurrence for general track-like conformal integrals, relating an integral in any even dimension to a two-dimensional integral with shifted propagator powers. This has as an immediate consequence that statements made about two-dimensional ladder integrals can be extended to higher (even) dimensions just by acting with a differential operator.

Let us therefore specialize to fully massless conformal ladder integrals \eqref{eq:defLadderIntegrals}, cf.\ \Figref{fig:ladders}. In the notation of the previous subsection this amounts to the following choice of index sets:
\begin{equation}
    S_1=\{ 1,2,4 \},\quad S_L=\{ 2,3,4 \},\quad S_i=\{ 2,4 \},\qquad  i=2,\dots,L-1.
\end{equation}
The reason for the recursion is now the fact that the squared distance $x_{13}^2$ only appears in the product $P_1\cdot P_L$ and thus (with Schwinger parameter $t_i$ for the propagator with weight $a_i$)
\begin{equation}
    I_{D;\bm{a}}^{(L)}=2^L\int[\dd t] \exp\left( -t_1\dots t_{L+1}x_{13}^2+(\text{independent of } x_{13}^2)  \right) \,,
\end{equation}
which immediately leads us to
\begin{equation}
    \frac{\partial}{\partial x_{13}^2}I_{D;\bm{a}}^{(L)}=-a_1\dots a_{L+1} I_{D+2;\bm{a}+\bm{e}_{1,\dots, L+1}}^{(L)}  \,,
\end{equation}
where $\mathbf{e}_{i_1,\dots,i_k}=\mathbf{e}_{i_1}+\dots +\mathbf{e}_{i_k}$ is a sum of standard unit vectors in the space of propagator powers. Note that the shifted propagator weights are precisely those along the horizontal and hence, in particular every vertex contains precisely two propagator weights that are shifted by 1, leading to a shift in the spacetime dimension of the integral by 2 due to the conformal constraints.

If we now, as in the previous subsection, strip off a kinematic factor carrying the conformal weights, the recursion takes a manifestly conformal form
\begin{equation}
   \frac{z\partial_z-\zb\partial_{\zb}}{z-\zb}\phi_{D;\bm{a}}^{(L)}(z,\zb)=-a_1\dots a_{L+1}\phi_{D+2;\bm{a}+\bm{e}_{1,\dots, L+1}}^{(L)}(z,\zb)\,.
\end{equation}
Such a recursion for ladder integrals has been observed before for a special choice of propagator powers, namely in the $\gamma\rightarrow 1$ limit of the choice given in eq.~\eqref{eq:propPowersOneParamDeformation} \cite{Petkou:2021zhg,Derkachov:2021ufp, Karydas:2023ufs}.

By iterating the recursion we can now express a generic conformal ladder integral in arbitrary even dimension as a derivative of the (shifted) two-dimensional result
\begin{equation}
    \label{eq:dDimLadderAs2DLadder}
    \phi^{(L)}_{D;\bm{a}}=\prod_{i=1}^{L+1}\frac{\Gamma(a_i-D/2+1)}{\Gamma(a_i)}\left[ \frac{z\partial_z-\zb \partial_{\zb}}{z-\zb}\right]^{D/2-1}\phi^{(L)}_{2;\bm{a}-(D/2-1)\bm{e}_{1,2,\dots,L+1}}\,.
\end{equation}
The 2d massless ladder integrals have recently been computed for generic propagator powers in terms of single-valued hypergeometric functions \cite{Duhr:2023bku}. These are single-valued bilinear combinations of ordinary hypergeometric functions, see \Secref{sec:2dDoubleCopy} for more details and references as well as \appref{sec:svHGFs} for an explicit definition. This immediately lets us write down the solution for the general conformal ladder integral in any even dimension as a derivative operator acting on a single-valued hypergeometric function
\begin{equation}
    \phi_{D;\bm{a}}^{(L)}(z,\zb)=\prod_{i=1}^{L+1}\frac{\Gamma(a_i-D/2+1)}{\Gamma(a_i)}\left[\frac{z\partial_z-\zb\partial_{\zb}}{z-\zb} \right]^{D/2-1}\psvFq{L+1}{L}{\bm{\alpha_a}}{\bm{\beta_a}}{z} \,.
    \label{eq:dimRecursion}
\end{equation}
Here the arguments of the hypergeometric function are related to the propagator powers by
\begin{align}
    (\bm{\alpha_a})_0&=1+a_{L+1}-h,\nonumber\\
    (\bm{\alpha_a})_k&=(L-k)h+1-a_{k+1,L}-a_{L+k+1,2L+1},~k=1,\dots,L, \\
    (\bm{\beta_a})_k&=(L-k+1)h+1-a_{k,L}-a_{L+k+1,2L+1},~k=1,\dots,L \,,\nonumber
\end{align}
where we recall that $h=D/2$ and we used the abbreviation $a_{i,j}=\sum_{k=i}^j a_k$.

While the unit-propagator limit of the above expression will be discussed in \Secref{subsec:SOVRep},
let us for now stick to generic propagator weights and illustrate the formula \eqref{eq:dimRecursion} in the simplest case of the box integral.


\paragraph{Example: Box Integral.}
For $L=1$ we find
\begin{equation}
    \phi_{D;\bm{a}}^{(1)}(z,\zb)=\frac{\Gamma_{a_1-h+1}\Gamma_{a_2-h+1}}{\Gamma_{a_1}\Gamma_{a_2}}  \left[\frac{1}{z-\zb}(z\partial_z-\zb\partial_{\zb}) \right]^{h-1} \psvFq{2}{1}{1-h+a_2,\, 1-a_3}{1+h-a_1-a_3}{z} \,,
\end{equation}
where the single-valued Gauss hypergeometric function is explicitly given by
\begin{align}
    & \psvFq{2}{1}{1-h+a_2,\,1-a_3}{1+h-a_1-a_3}{z}= \frac{\Gamma_{h-a_1}\Gamma_{1-a_3}\Gamma_{a_1+a_3-h}}{\Gamma_{1-h+a_1}\Gamma_{1+h-a_1-a_3}\Gamma_{a_3}}\left|\pFq{2}{1}{1-h+a_2,\, 1-a_3}{1+h-a_1-a_3}{z}\right|^2 \\
    &\quad -\frac{\Gamma_{h-a_2}\Gamma_{1-D+a_1+a_2+a_3}\Gamma_{a_1+a_3-h}\Gamma_{1+h-a_1-a_3}}{\Gamma_{1-h+a_2}\Gamma_{1-h+a_1+a_3}^2\Gamma_{D-a_1-a_2-a_3}}  (z\zb)^{a_1+a_3-h}\left|\pFq{2}{1}{1-h+a_1,\,1-D+a_1+a_2+a_3}{1-h+a_1+a_3}{z}\right|^2 \,, \nonumber
    \label{eq:2F1sv}
\end{align}
and we use the abbreviation $\Gamma_x=\Gamma(x)$. It hence follows that the conformal box integral in any even dimension can be written as a bilinear in Gauss hypergeometric functions (since the derivatives only lead to shifts in the arguments). 

While the above form is a very natural hypergeometric expression in terms of the variables $z,\zb$, we can also find a simple expression in terms of hypergeometric functions of the conformal cross ratios $u,v$ defined by
\begin{equation}
    u=z\zb,\qquad v=(1-z)(1-\zb) \, .
\end{equation}
This can be achieved by using the well-known transformation formulas for the Gauss hypergeometric function
\begin{align}
    \pFq{2}{1}{\alpha, \, \beta}{\gamma}{x} &= \frac{\Gamma_\gamma\Gamma_{\gamma-\alpha-\beta}}{\Gamma_{\gamma-\alpha}\Gamma_{\gamma-\beta}}
    \pFq{2}{1}{\alpha, \, \beta}{\alpha+\beta+1-\gamma}{1-x} \nonumber\\
    &\quad +\frac{\Gamma_\gamma\Gamma_{\alpha+\beta-\gamma}}{\Gamma_{\alpha}\Gamma_{\beta}} (1-x)^{\gamma-\alpha-\beta}
    \pFq{2}{1}{\gamma-\alpha, \, \gamma-\beta}{1+\gamma-\alpha-\beta}{1-x} \, ,\\
    \pFq{2}{1}{\alpha, \, \beta}{\gamma}{x} &= (1-x)^{\gamma-\alpha-\beta} \pFq{2}{1}{\gamma-\alpha,\,\gamma-\beta}{\gamma}{x} \, ,
\end{align}
as well as the relation \cite{f4Expansion1,f4Expansion2}
\begin{equation}
    \FF{4}{\alpha, \, \beta}{\gamma, \, \gamma'}{u,v}=\pFq{2}{1}{\alpha, \, \beta}{\gamma}{z} \pFq{2}{1}{\alpha, \, \beta}{\gamma'}{1-\zb} \, ,
\end{equation}
which holds for $\gamma+\gamma'=\alpha+\beta+1$. This yields an expression for the two-dimensional box integral in terms of Appell $F_4$ functions in the conformal cross ratios $u,v$. This expression precisely agrees with the well-known expression of the $D$-dimensional conformal box integral, see e.g.~\cite{Loebbert:2019vcj}, specialized to $D=2$.

We can also rediscover the general expression in other even dimensions using the recursion. In the $u,v$ variables the recursion operator reads
\begin{equation}
    \frac{z\partial_z-\zb\partial_{\zb}}{z-\zb}=-\frac{\partial}{\partial v}.
\end{equation}
Noticing that the derivatives only shift the arguments of the $F_4$ functions
\begin{equation}
    \frac{\partial}{\partial v} \FF{4}{\alpha, \, \beta}{\gamma, \, \gamma'}{u,v} =\frac{\alpha\beta}{\gamma'}\FF{4}{\alpha+1, \, \beta+1}{\gamma, \, \gamma'+1}{u,v} \, ,
\end{equation}
allows us to reproduce the $D$-dimensional result of the form as in \cite{Loebbert:2019vcj} for any even dimension after using contiguous relations for Appell $F_4$ functions, see for example \cite{f4Relations}.

\subsection{Example: Lifting Calabi--Yaus from Two to Higher Dimensions}
\label{subsec:CYlift}
In this subsection we want to focus on a particular connection between two-dimensional integrals with certain rational propagator powers and Calabi--Yau varieties, as discussed in \cite{Duhr:2022pch,Duhr:2023eld,Duhr:2024hjf}.
 
As conjectured in \cite{Duhr:2022pch} and explicitly shown in \cite{Duhr:2023eld}, two-dimensional ladder integrals with all propagator weights equal to $1/2$ can be written as bilinears in periods of certain Calabi--Yau varieties
\begin{equation}
    \phi_{2;\gamma=1/2}^{(L)}(z,\bar{z})=\Pi_L(z)^\dagger\Sigma_L\Pi_L(z)\,.
\end{equation}
Here $\Pi_L(z)$ is the period vector and $\Sigma$ is the so-called intersection matrix and is chosen in such a way that the result is single-valued. This works in the same way as for single-valued hypergeometric functions as briefly explained in \appref{sec:svHGFs}. We refer to \cite{Duhr:2022pch,Duhr:2023eld,Duhr:2024hjf} for more details.

As a simple example consider the box integral. The periods can be found by solving the so-called Picard-Fuchs equation, which is explicitly given by \cite{Duhr:2023eld}
\begin{equation}
    \left[\theta^{2}-z\left(\theta+\frac{1}{2} \right)^{2}\right]\Pi_1(z)=0\,,\qquad \theta=z\partial_z\,.
\end{equation}
A basis for the solution space can be written in terms of the complete elliptic integral of the first kind
\begin{equation}
    \Pi_1(z)=\left( K(z),iK(1-z) \right) \,.
\end{equation}
These are periods of an elliptic curve, i.e.~a Calabi--Yau 1-fold. The intersection form, i.e.~the bilinear form defined by the intersection matrix
\begin{equation}
    \Sigma_1=\begin{pmatrix}
        0 & 1 \\ -1 & 0
    \end{pmatrix} \,
\end{equation}
is invariant under all matrices in the monodromy group of $\Pi_1$ and leads to the expression (the prefactor can e.g.\ be fixed by numerical evaluation at a certain point)
\begin{equation}
    \phi_{2;\gamma=1/2}^{(1)}(z,\zb)=\frac{4}{i\pi}\Pi_1(z)^\dagger\Sigma_1\Pi_1(z) \,.
\end{equation}
This result has also been computed before in \cite{Derkachov:2018rot,Corcoran:2021gda}, without reference to the underlying Calabi--Yau geometry.

A direct consequence of the recursion from the previous subsection is now that we can write the higher dimensional ladder integrals with propagator powers $(D-1)/2$ on the horizontal and $1/2$ on the vertical as a bilinear in (derivatives of) the periods of the Calabi--Yau variety. For example the four-dimensional ladder integrals (with propagator powers 3/2 on the horizontal and 1/2 on the vertical) can be written as
\begin{align}
    \phi_{4;\gamma=1/2}^{(L)}(z,\bar{z})&=\frac{2^{L+1}}{z-\bar{z}}(\theta_z-\theta_{\bar{z}})\left[ \Pi_L(z)^\dagger\Sigma_L\Pi_L(z)\right] \nonumber \\
    &=\frac{2^{L+1}}{z-\bar{z}}
    \begin{pmatrix}
        \Pi_L(z) \\ \theta_z\Pi_L(z)
    \end{pmatrix}^\dagger
    \begin{pmatrix}
        0 & \Sigma_L \\ -\Sigma_L & 0
    \end{pmatrix}
    \begin{pmatrix}
        \Pi_L(z) \\ \theta_z\Pi_L(z)
    \end{pmatrix}.
\end{align}
It turns out that the fact that conformal ladder integrals can be written in a \enquote{double copy form} is actually more general in both two dimensions and, as a consequence of the recursion, also in higher dimensions. This furthermore extends to other four-point integrals to which the recursion applies, as well as to higher-point integrals restricted to two-dimensional kinematics. We will discuss this in more detail in \Secref{sec:doubleCopy}.

The fact that some Feynman integrals evaluate to (integrals over) periods of Calabi--Yau varieties has been known for a couple of years and families of these integrals have received a considerable amount of attention since, see \cite{Bourjaily:2022bwx} for a review. Using the dimensional recursion we can now easily extend the list of higher-dimensional integrals connected to Calabi--Yau varieties. We have already mentioned the example of the ladder integrals. Using the results of \cite{Duhr:2024hjf} we can also add higher point examples as long as we restrict to two-dimensional kinematics, such as the train track integrals as well as the triangle track integrals (see \cite{Duhr:2024hjf} for more details).
\subsection{Relation to Standard Dimension Shifts, Conformal Blocks and Witten Diagrams}
Let us comment on interesting connections to related dimensional recursion relations considered in other contexts. 
\paragraph{Dimension Shift Relations for Feynman Integrals.}
Let us briefly sketch in which sense the above dimension shift relations are advantageous over the traditional dimension shifts for Feynman integrals \cite{TaraSOV:1996br,TaraSOV:1997kx,Lee:2009dh}. These take the form\footnote{We are ignoring some subtleties here about possible auxiliary propagators that might need to be added since they do not change the argument. We refer to the original references or the review \cite{Abreu:2022mfk} for more details.}
\begin{equation}
    I_{D+2;\bm{a}}=\frac{2^L\Gamma(D-L-n+2) G(x_{12},x_{23},\dots,x_{n-1,n})}{\Gamma(D-n+2)}\mathcal{B}(b_1,\dots,b_{L+n-1})I_{D;\bm{a}}\,,
\end{equation}
for some $L$-loop Feynman integral depending on $n$ external points. Here $G(k_1,\dots,k_m)=\det(k_i\cdot k_j)$ is a Gram determinant and $\mathcal{B}(z_1,\dots,z_K)$ is the so-called Baikov polynomial of the integral \cite{Baikov:1996iu}, whose precise definition does not play a role for us. The arguments of the Baikov polynomial are operators shifting the propagator weights of the Feynman integral
\begin{equation}
    b_i I_{D;\bm{a}}=I_{D;\bm{a}-\bm{e}_i}\,.
\end{equation}
Hence it is clear that we can compute a Feynman integral in $D+2$ dimensions as a linear combination of $D$-dimensional Feynman integrals with shifted propagator powers. Since Feynman integrals with a fixed set of propagators form a finite-dimensional vector space \cite{Smirnov:2010hn} one can use integration-by-parts (IBP) identities \cite{Tkachov:1981wb,Chetyrkin:1981qh} to write the integrals appearing on the right hand side in terms of some basis.

Since differentiating a Feynman integral also has the effect of shifting the propagator powers we could also express the result of the action of our dimension shift operator in terms of the same basis of Feynman integrals and hence rediscover the familiar dimension shift relation. However, the above dimension shift relation for ladder integrals here is clearly simpler as it only needs the knowledge of a single integral in the lower dimension and furthermore is manifestly conformally invariant. 

\paragraph{Witten Diagrams.}

Witten diagrams \cite{Witten:1998qj} are the analogues of Feynman diagrams in Anti-de Sitter (AdS) space. The simplest family of Witten diagrams is given by contact diagrams, defined as
\begin{equation}
    W^{d;\text{contact}}_{\bm{\Delta}}(x_1,\dots,x_n)=\int_0^\infty\frac{\dd z_0}{z_0^d}\int_{\mathbb{R}^d}\dd^d z\prod_{i=1}^n\left( \frac{z_0}{z_0^2+(z-x_i)^2}\right)^{\Delta_i},
\end{equation}
were $(z_0,z)\in\mathbb{R}_{\geq 0}\times\mathbb{R}^d$ are Poincar\'e coordinates for the $\mathrm{AdS}_{d+1}$ space, where $d$ is the boundary dimension, $x_1,\dots, x_n$ are points on the conformal boundary and the $\Delta_i$ represent the external scaling dimensions. The latter can be interpreted as the scaling dimensions of operators in the dual CFT that are sourced by the bulk fields propagating in the Witten diagram. From the bulk perspective they are related to the masses of the bulk fields, see e.g.~the review \cite{DHoker:2002nbb} for details.

It has been known since the early days of the AdS/CFT correspondence \cite{DHoker:1999kzh}, that these contact Witten diagrams satisfy recursion relations (see also \cite{Rigatos:2022eos}):
\begin{equation}
    \frac{\partial}{\partial x_{ij}^2}W^{d;\text{contact}}_{\bm{\Delta}}(x_1,\dots,x_n)=\frac{2\Delta_i\Delta_j}{d-1-\sum_{i=1}^n\Delta_i}W^{d;\text{contact}}_{\bm{\Delta}+\bm{e}_{i,j}}(x_1,\dots,x_n).
\end{equation}

These are however not recursion relations in the boundary dimension but only in the external scaling dimensions, since there is no relation between the $\Delta_i$ and $d$. In turns out that contact diagrams are actually independent of the boundary dimensions up to a prefactor \cite{Rastelli:2017ecj,Rigatos:2022eos}.

The recursion relations for contact Witten diagrams can also be shown to follow from a conformal Schwinger parametrization \cite{Rastelli:2017ecj,Rigatos:2022eos}. This parametrization  implies a direct relation between contact Witten diagrams and conformal Feynman integrals
\begin{equation}
    W^{d;\text{contact}}_{\bm{\Delta}}(x_1,\dots,x_n)=\frac{1}{2}\pi^{\frac{d-1}{2}}\Gamma\left(\frac{\sum_{i=1}^n\Delta_i-d+1}{2} \right)I_{D;\bm{\Delta}}^{(1)}(x_1,\dots,x_n),
\end{equation}
where the dimension of the Feynman integral is given by $D=\sum_{i=1}^n\Delta_i$. This makes it clear that the recursion for conformal Feynman integrals at one loop found here exactly coincides with the recursion for contact Witten diagrams. Since here we have generalized the recursion to arbitrary track-like Feynman integrals (with the mentioned subtleties beyond four points), it is a natural question to ask if similar recursions also hold for Witten diagrams beyond the contact case. We leave this for future investigation

\paragraph{Conformal Blocks.}

Finally, let us comment on the implications of the dimensional recursion for conformal blocks, the building blocks of the conformal block expansion which is an important tool in any conformal field theory. For example a conformal correlation function of four scalar operators with scaling dimensions $\Delta_1,\dots,\Delta_4$ can be expanded as
\begin{equation}
    \langle \mathcal{O}_{\Delta_1}(x_1)\mathcal{O}_{\Delta_2}(x_2)\mathcal{O}_{\Delta_3}(x_3)\mathcal{O}_{\Delta_4}(x_4) \rangle=V_{\bm{\Delta}}(x_1,\dots,x_4)\sum_{\mathcal{O}_{\Delta,l}\text{ primary}}\lambda_{\mathcal{O}_1,\mathcal{O}_2,\mathcal{O}}\lambda_{\mathcal{O}_3,\mathcal{O}_4,\mathcal{O}}g_{\Delta,l,D}(z,\zb),
\end{equation}
where $V_{\bm{\Delta}}$ is a function of the external points and scaling dimensions carrying the conformal weights and the $\lambda_{\mathcal{O}_i,\mathcal{O}_j,\mathcal{O}_k}$ are the conformal structure constants, which are determined by the three-point correlation function of the operators $\mathcal{O}_i,\mathcal{O}_j,\mathcal{O}_k$. The sum is taken over all primary operators labelled by their scaling dimensions $\Delta$ and spin $l$. Both the structure constants as well as the details of the summation are theory dependent, while the conformal blocks $g_{\Delta,l,D}(z,\zb)$ are fully determined by conformal symmetry and hence universal.

While conformal block expansions have been known for a long time \cite{Ferrara:1972kab}, they have really become useful due to the seminal papers of Dolan and Osborn \cite{Dolan:2000ut,Dolan:2003hv}, see also \cite{Dolan:2011dv}, in which they characterised conformal blocks as eigenfunctions of the conformal Casimir operator and in particular found simple hypergeometric expressions for four-point conformal blocks. The advancement in the understanding of conformal blocks has in particular lead to the development of the conformal bootstrap program, which crucially builds on efficient ways of numerically evaluating conformal blocks, see e.g.~\cite{Poland:2018epd,Hartman:2022zik,Poland:2022qrs} for reviews.

In \cite{Simmons-Duffin:2012juh} it was shown that conformal blocks can be expressed as (monodromy-projected) conformal Feynman integrals. Explicitly, consider the conformal box integral, rescaled as follows
\begin{equation}
    F_{\Delta,D}(z,\zb)=\frac{\left(\frac{\Delta_3-\Delta_4+\Delta}{2} \right)_{h-\Delta}\left(\frac{\Delta_4-\Delta_3+\Delta}{2} \right)_{h-\Delta}}{(\Delta)_{h-2\Delta}}(z\zb)^{h-\frac{\Delta}{2}}((1-z)(1-\zb))^{\frac{\Delta_3-\Delta_4}{2}}\phi_{D;\bm{a}_{\Delta}}^{(1)}(z,\zb) \,,
\end{equation}
where $(x)_n$ denotes the Pochhammer symbol and the propagator powers are related to the external scaling dimensions of the conformal correlation function as
\begin{equation}
    \bm{a}_\Delta=\left(\frac{\Delta_1-\Delta_2+\Delta}{2},\frac{D+\Delta_3-\Delta_4-\Delta}{2},\frac{\Delta_2-\Delta_1+\Delta}{2},\frac{D-\Delta_3+\Delta_4-\Delta}{2} \right).
\end{equation}
This object is usually referred to as \textit{conformal partial wave} and can be expressed in terms of a conformal block $g$ and its so-called \textit{shadow block} $\tilde g$:
\begin{equation}
\label{eq:shadowBlockDecomposition}
    F_{\Delta,D}(z,\zb)=g_{\Delta,0,D}(z,\zb)+K_{\Delta,D}\tilde{g}_{D-\Delta,0,D}(z,\zb)\,.
\end{equation}
Here $K_{\Delta,D}$ is some constant ratio of Gamma functions ensuring that the conformal partial wave is single-valued. The shadow block \cite{Ferrara:1972uq} $\tilde{g}_{D-\Delta,D}(z,\zb)$ is also an eigenfunction of the same Casimir operator but it has the unphysical scaling dimension $D-\Delta$. The two can for example be distinguished by their behaviour under the monodromy action $x_{12}^2\rightarrow e^{4\pi i} x_{12}^2$, which can be used to remove the shadow block contribution by projecting onto a certain eigenspace of the monodromy, see \cite{Simmons-Duffin:2012juh} for details.

The recursion that we discussed for conformal Feynman integrals now directly implies a recursion relation for the four-point scalar partial wave and hence through eq.~\eqref{eq:shadowBlockDecomposition}
for the associated conformal block. Indeed in \cite{Simmons-Duffin:2012juh} the four-point block in general even dimension was computed as a derivative of the two-dimensional result.

Also higher point conformal partial waves can be written as conformal Feynman integrals as explained in \cite{Simmons-Duffin:2012juh}. Hence we should also be able to make use of the recursion here and find the conformal blocks in higher dimensions as derivatives of the two-dimensional result, at least in two-dimensional kinematics. Indeed the comb channel conformal blocks in two dimensions have recently been computed for any number of external points \cite{Rosenhaus:2018zqn}, see also \cite{Alkalaev:2015fbw,Bhatta:2016hpz,Besken:2016ooo,Fortin:2020zxw,Alkalaev:2023axo,Fortin:2023xqq} for related work on one-and two-dimensional conformal blocks. We will leave this for future investigation, see also \cite{Hoback:2020syd} for related work.


\subsection{Relation to Separation of Variables Representation}
\label{subsec:SOVRep}

In this subsection we briefly connect the above observations to the separation of variables representation of fishnet integrals, which will be useful in the subsequent sections.

\paragraph{Harmonic Analysis.}
As announced in the introduction, there are two differential operators which play a key role in this paper, namely\footnote{These two operators are (up to prefactors) precisely the Cartan generators of the two-dimensional conformal algebra.}
\begin{equation}
\dimOp=\frac{z\partial_z-\zb\partial_{\zb}}{z-\zb} \,,
\qquad
\loopOp=-\frac{z\partial_z+\zb\partial_{\zb}}{\log(z\zb)} \,.
\end{equation}
We have already seen that the first operator induces a dimensional recursion on track-like integrals. In the next section we will further study the action of the second operator.

Let us now, however, first discuss a representation of conformal Feynman integrals which trivializes the action of these two operators and hence provides a natural arena to study them. This representation is given by expanding into the eigenfunctions of these two operators which, at least formally, is easily done by realising that these two operators are precisely proportional to an angular and radial (logarithmic) derivative
\begin{equation}
    \dimOp\sim\frac{\partial}{\partial\varphi},\qquad \loopOp\sim r\frac{\partial}{\partial r} \,,
\end{equation}
after parametrizing the complex $z$ plane as
\begin{equation}
    z=re^{i\varphi},\qquad \zb=r e^{-i\varphi} \,.
\end{equation}
Expanding in the eigenfunctions of the two operators now corresponds to
\begin{align*}
    \phi(z,\zb)&=\sum_{n=-\infty}^\infty\int_{-\infty}^\infty\frac{\dd\nu}{2\pi}\phi(n,\nu)e^{2i\nu\log r}e^{i n\varphi} \\
    &=\sum_{n=-\infty}^\infty\int_{-\infty}^\infty\frac{\dd\nu}{2\pi}\phi(n,\nu)(z\zb)^{i\nu}\left(\frac{z}{\zb}\right)^{n/2} \,.
\end{align*}
This representation can be viewed as a \enquote{conformal Fourier transformation} and has been called Fourier--Mellin representation in the literature \cite{fourierMellin,Dixon:2012yy}.


\paragraph{Fourier--Mellin versus Separation of Variables Representation. }
In general, the explicit form of the Fourier--Mellin representation, i.e.\ the form of the coefficients $\phi(n,\nu)$, is difficult to derive but for ladder integrals for a special choice of propagator powers this has been achieved \cite{Fleury:2016ykk,Derkachov:2018rot,Derkachov:2019tzo,Derkachov:2020zvv,Basso:2021omx,Derkachov:2021ufp}. This choice amounts to taking the vertical propagators to have weights $\gamma$ and the horizontal propagators to have weights $D/2-\gamma$ with $\gamma\in (0,D/2)$. For the ladder integrals this means explicitly that we choose
\begin{align}
    a_i&=\frac{D}{2}-\gamma, \qquad i=1,\dots,L+1 \,, 
    \nonumber\\
    a_i&=\gamma, \qquad i=L+2,\dots, 3L+1 \,.
    \label{eq:propPowersOneParamDeformation}
\end{align}
Note that this in particular includes the unit-propagator power case in four dimensions. The above choice might seem  ad hoc, but it has a natural interpretation from the perspective of integrability. In fact, the ladder integrals (or four-point fishnet integrals in general) with this choice of propagator powers are precisely computing certain four-point correlation functions in the deformed fishnet CFT proposed in \cite{Kazakov:2018qbr}, which is a $D$-dimensional generalization of the four-dimensional fishnet CFT of \cite{Gurdogan:2015csr}. These deformation parameters can furthermore be interpreted as angles in the Baxter lattice picture of \cite{Zamolodchikov:1980mb,Kazakov:2022dbd}.

The Fourier--Mellin representation of the ladder integrals for this choice of propagator powers is explicitly given in 2d by\cite{Derkachov:2018rot}
\begin{align}
\label{eq:2dLaddersFourierMellin}
    \phi_{2;\gamma}^{(L)}&(z,\zb)=(z\zb)^{\frac{\gamma-1}{2}}\left( \frac{\Gamma(\gamma)}{\Gamma(1-\gamma)} \right)^{L+1}\\
    &\times\sum_{n=-\infty}^\infty\int_{-\infty}^\infty\frac{\dd\nu}{2\pi}\left((-1)^n\frac{\Gamma\left(\frac{1-\gamma}{2}+\frac{n}{2}-i\nu\right)\Gamma\left(\frac{1-\gamma}{2}-\frac{n}{2}+i\nu\right)}{\Gamma\left(\frac{\gamma+1}{2}+\frac{n}{2}+i\nu\right)\Gamma\left(\frac{\gamma+1}{2}-\frac{n}{2}-i\nu\right)} \right)^{L+1}(z\zb)^{i\nu}\left(\frac{z}{\zb} \right)^{\frac{n}{2}} \,, \nonumber
\end{align}
as well as in 4d by\footnote{In the undeformed limit $\gamma\rightarrow 1$ this representation has been computed before using hexagonalization \cite{Fleury:2016ykk}.} \cite{Derkachov:2019tzo,Derkachov:2020zvv,Basso:2021omx,Derkachov:2021ufp}
\begin{align}
\label{eq:4dLaddersFourierMellin}
    \phi_{4;\gamma}^{(L)}&(z,\zb)=\frac{(z\zb)^{\frac{\gamma-1}{2}}}{z-\zb}\left( \frac{\Gamma(\gamma)}{\Gamma(2-\gamma)} \right)^{L+1}\\
    &\times\sum_{n=-\infty}^\infty\int_{-\infty}^\infty\frac{\dd\nu}{2\pi}n\left((-1)^n\frac{\Gamma\left(\frac{1-\gamma}{2}+\frac{n}{2}-i\nu\right)\Gamma\left(\frac{1-\gamma}{2}-\frac{n}{2}+i\nu\right)}{\Gamma\left(\frac{\gamma+1}{2}+\frac{n}{2}+i\nu\right)\Gamma\left(\frac{\gamma+1}{2}-\frac{n}{2}-i\nu\right)} \right)^{L+1}(z\zb)^{i\nu}\left(\frac{z}{\zb} \right)^{\frac{n}{2}} \,.
    \nonumber
\end{align}
Indeed, since the dimensional recursion operator acts as multiplication by $n$ in Fourier--Mellin space, we can see that the recursion is made completely manifest in this representation. 

We could now also study the action of the other operator $\dimOp$ in the Fourier--Mellin representation, and we will do so in the next section, but let us first mention that the Fourier--Mellin representation of the ladder integrals given above is actually a special case of the so-called \textit{separation of variables (SOV) representation} based on the SOV method of integrability \cite{Sklyanin:1987ih,Sklyanin:1995bm}. We will use the rest of this section to review this representation and illustrate its use.

\begin{figure}[t]
\begin{center}
\includegraphicsbox[]{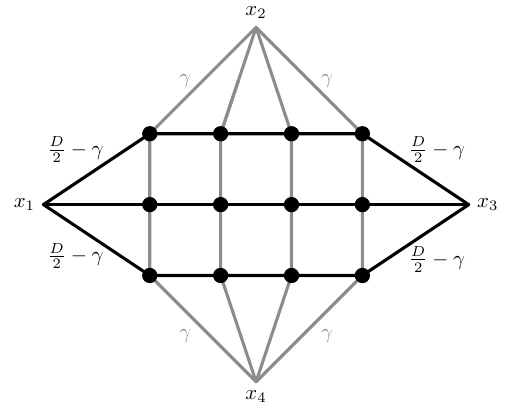}
\end{center}
\caption{Example of a four-point (or Basso--Dixon) fishnet integral with $M=3,N=4$. Here, horizonal or vertical propagator powers take values $D/2-\gamma$ or $\gamma$, respectively. The ladder integrals ($M=1$) with a parameter $\gamma$ correspond to the situation of a single row of integration vertices.}
\label{fig:fourpointfishnetgamma}
\end{figure}

The SOV representation has been derived for four-point fishnet integrals (see \Figref{fig:fourpointfishnetgamma}) with the above choice of propagator powers both in two dimensions \cite{Derkachov:2018rot} 
\begin{align}
    I_{2;\gamma}^{(M,N)}&=\frac{\left(x_{14}^2\right)^{(\gamma-1)M}\left(x_{23}^2\right)^{(\gamma-1)M}}{\left(x_{24}^2\right)^{(\gamma-1)M+\gamma N}}\frac{(z\zb)^{\frac{\gamma-1}{2}M}}{M!}\left(\frac{\Gamma(\gamma)}{\Gamma(1-\gamma)}\right)^{M(M+N)} \nonumber \\
    &\quad \times \left( \prod_{i=1}^M\sum_{n_i=-\infty}^\infty\int_{-\infty}^\infty\frac{\dd\nu_i}{2\pi}\right) \prod_{j<k=1}^M\left[ \frac{(n_j-n_k)^2}{4}+(\nu_j-\nu_k)^2\right]  \label{eq:2dSOV} \\
    &\quad \times\prod_{k=1}^M\left[ (-1)^{n_k}\frac{\Gamma\left(\frac{1-\gamma}{2}+\frac{n_k}{2}-i\nu_k \right)\Gamma\left(\frac{1-\gamma}{2}-\frac{n_k}{2}+i\nu_k \right)}{\Gamma\left(\frac{1+\gamma}{2}-\frac{n_k}{2}-i\nu_k \right)\Gamma\left(\frac{1+\gamma}{2}+\frac{n_k}{2}+i\nu_k \right)}\right]^{M+N}(z\zb)^{i\nu_k}\left( \frac{z}{\zb}\right)^{\frac{n_k}{2}} \nonumber \, ,
\end{align}
as well as in four dimensions (first for $\gamma=1$ \cite{Basso:2017jwq} and then for general $\gamma$ \cite{Derkachov:2019tzo,Derkachov:2020zvv,Basso:2021omx,Derkachov:2021ufp})
\begin{align}
    I_{4;\gamma}^{(M,N)}&=\frac{\left(x_{14}^2\right)^{(\gamma-2)M}\left(x_{23}^2\right)^{(\gamma-2)M}}{\left(x_{24}^2\right)^{(\gamma-2)M+\gamma N}}\frac{(z\zb)^{\frac{\gamma-1}{2}M}}{(z-\zb)^M}\frac{1}{M!}\left( \frac{\Gamma(\gamma)}{\Gamma(2-\gamma)}\right)^{M(M+N)} \label{eq:4dSOV} \\
    &\quad \times\left( \prod_{i=1}^M\sum_{n_i=-\infty}^\infty\int_{-\infty}^\infty\frac{\dd\nu_i}{2\pi}\right)\prod_{j<k=1}^M\left[ \frac{(n_j-n_k)^2}{4}+(\nu_j-\nu_k)^2\right]\left[ \frac{(n_j+n_k)^2}{4}+(\nu_j-\nu_k)^2\right] \nonumber\\
    &\quad \times \prod_{k=1}^M a_k\left[(-1)^{n_k}\frac{\Gamma\left(\frac{1-\gamma}{2}+\frac{n_k}{2}-i\nu_k \right)\Gamma\left(\frac{1-\gamma}{2}-\frac{n_k}{2}+i\nu_k \right)}{\Gamma\left(\frac{1+\gamma}{2}-\frac{n_k}{2}-i\nu_k \right)\Gamma\left(\frac{1+\gamma}{2}+\frac{n_k}{2}+i\nu_k \right)}\right]^{M+N}(z\zb)^{i\nu_k}\left( \frac{z}{\zb}\right)^{\frac{n_k}{2}}  \,. \nonumber
\end{align}
This representation has also been derived for fishnet theories in general dimensions \cite{Derkachov:2021ufp}, involving fermions \cite{Derkachov:2020zvv} or with more general choices of propagator weights such as the checkerboard CFT \cite{Alfimov:2023vev}.

For $M=1$, i.e.~for the ladder integrals, this representation is precisely the Fourier--Mellin representation given in eqs.~\eqref{eq:2dLaddersFourierMellin} and \eqref{eq:4dLaddersFourierMellin}. 
For $M>1$, i.e.\ for more \enquote{layers} in the diagram, the representation involves multiple sums and integrals and hence in particular provides an integral/series representation for these integrals in Fourier--Mellin space. More explicitly the SOV representation can be written as (up to some prefactors)
\begin{equation}
    \left( \prod_{i=1}^M\sum_{n_i=-\infty}^\infty\int_{-\infty}^\infty\frac{\dd\nu_i}{2\pi}\right)\mathcal{I}^{(\gamma)}_{M,N}(\bm{n},\bm{\nu})(z\zb)^{i\sum_{k=1}^M\nu_k}\left(\frac{z}{\zb} \right)^{\frac{1}{2}\sum_{k=1}^M n_k} \,.
\end{equation}
Here the integrand $\mathcal{I}^{(\gamma)}_{M,N}(\bm{n},\bm{\nu})$ depends on the summation and integration variables $\bm{n}=(n_1,\dots,n_M)$ and $\bm{\nu}=(\nu_1,\dots,\nu_M)$ and can be read off from the explicit formulas \eqref{eq:2dSOV,eq:4dSOV}. By changing variables to $(n_+,n_1,\dots ,n_{M-1})$ and $(\nu_+,\nu_1,\dots ,\nu_{M-1})$ (or similarly), where $n_+=\sum_{k=1}^M n_k,\nu_+=\sum_{k=1}^M\nu_k$ we see that the SOV representation yields an integral/series representation for the Fourier--Mellin representation of the integral of interest
\begin{equation}
    \sum_{n_+=-\infty}^\infty\int_{-\infty}^\infty\frac{\dd\nu_+}{2\pi}\left[\left( \prod_{i=1}^{M-1}\sum_{n_i=-\infty}^\infty\int_{-\infty}^\infty\frac{\dd\nu_i}{2\pi}\right)\mathcal{I}^{(\gamma)}_{M,N}(\bm{n},\bm{\nu}) \right](z\zb)^{i\nu_+}\left(\frac{z}{\zb} \right)^{\frac{n_+}{2}} \,.
\end{equation}
While in principle the integrals and sums could be carried out, this can be technically quite challenging in practice.

\paragraph{Polylogs and Unit-Propagator Limit of Ladder Integrals.}

Let us illustrate the use of the SOV (or Fourier--Mellin) representation to make sense of the dimensional recursion for ladder integrals in the unit-propagator limit, which will lead us to the famous formula \cite{Usyukina:1992jd,Usyukina:1993ch} for the four-dimensional unit-propagator weight ladder integrals in terms of classical polylogarithms from the above recursion \eqref{eq:dDimLadderAs2DLadder}. At the same time this allows to derive other all-loop formulas for polylogarithmic ladder integrals in other even dimensions (albeit with somewhat unusual choices of propagator weights).

Indeed the unit-propagator power limit is quite subtle, since the formula \eqref{eq:dDimLadderAs2DLadder} does not seem applicable with the Gamma-functions in the prefactor diverging for $a_i=1,h=2$. This divergence reflects the fact that the two-dimensional \enquote{seed} of the recursion is given by the two-dimensional ladder integrals with propagator weights zero on the horizontal, which are singular. A solution to this conundrum was first proposed in \cite{Karydas:2023ufs} and we will review it below. 

As a first step we take the propagator powers according to eq.~\eqref{eq:propPowersOneParamDeformation}, leading us to
\begin{equation}
    \phi_{4;\gamma}^{(L)}(z,\zb)=\left( \frac{\Gamma(1-\gamma)}{\Gamma(2-\gamma)} \right)^{L+1}\left[\frac{1}{z-\zb}(z\partial_z-\zb\partial_{\zb}) \right]\phi_{2;\gamma}^{(L)}(z,\zb) \,.
    \label{eq:dimRecursionOneParamDef}
\end{equation}
The divergence in the unit-propagator power limit $\gamma\rightarrow 1$ is now manifested by the $\Gamma(1-\gamma)$ in the prefactor.

In principle the four-dimensional ladder integral is still well-defined by the recursion also in the limit $\gamma\rightarrow 1$, since we can expand the two-dimensional ladder integral, expressed in terms of a single-valued hypergeometric function, in $\gamma-1$ and find that the coefficients of all negative powers lie in the kernel of the differential operator inducing the dimensional recursion. Hence we can simply drop them and only keep the finite part in the limit giving the four-dimensional ladder in terms of (parameter-derivatives of) hypergeometric functions. The expressions found in this way are, however, not very illuminating and in particular it is not so straightforward\footnote{Tools like \cite{Huber:2005yg} would be very helpful here.} to analytically match the simple polylogarithmic formulas of \cite{Usyukina:1992jd,Usyukina:1993ch}.

There is however a very nice alternative way of making sense of the $\gamma\rightarrow 1$ limit, first proposed in \cite{Karydas:2023ufs}, that gives a meaning to the two-dimensional ladders in this limit and easily reproduces the polylogarithmic results for the ladders. The key is the SOV/Fourier--Mellin representation for the two-dimensional ladder integrals \eqref{eq:2dLaddersFourierMellin}. Indeed, the prefactor in eq.~\eqref{eq:2dLaddersFourierMellin} precisely cancels the divergence in the dimensional recursion. Hence we can just define the regularised two-dimensional ladders in this limit by dropping the prefactor leading to the simple recursion
\begin{equation}
    \phi_{4;\gamma=1}^{(L)}(z,\zb)=\frac{z\partial_z-\zb\partial_{\zb}}{z-\zb}\phi_{2;\gamma=1}^{\text{reg},(L)}(z,\zb) \,,
\end{equation}
with\footnote{Note that the $n=0$ contribution is somewhat subtle. One can either understand it as a Cauchy principal value \cite{Karydas:2023ufs} or deform the contour to find a non-zero residue and then divide by 2 \cite{Dixon:2012yy}.}
\begin{equation}
    \phi_{2;\gamma=1}^{\text{reg},(L)}(z,\zb)=\sum_{n=-\infty}^\infty\int_{-\infty}^{\infty}\frac{\dd\nu}{2\pi}\frac{1}{\left(\nu^2+\frac{n^2}{4} \right)^{L+1}}(z\zb)^{i\nu}\left( \frac{z}{\zb}\right)^{n/2}.
\end{equation}
Indeed, since the differential operator just pulls down a factor of $n$ into the integral this naturally reproduces the Fourier--Mellin representation for the ladder integrals with unit propagator powers, derived for example in \cite{Fleury:2016ykk,Basso:2021omx} using integrability-based methods.

The regularised two-dimensional ladders can be explitly evaluated in terms of classical polylogarithms using the residue theorem, leading to \cite{Karydas:2023ufs}
\begin{align}
    \phi_{2;\gamma=1}^{\text{reg},(L)}(z,\zb)&=\sum_{n=0}^L\frac{(-1)^n(2L-n)!}{L!(L-n)! n!}\log(z\zb)^n(\Li_{2L+1-n}(z)+\Li_{2L+1-n}(\zb)) \nonumber \\
    &\qquad\qquad -\frac{(-1)^L}{2(2L+1)!}\log(z\zb)^{2L+1} \,.
\end{align}
It can easily be checked that this indeed reproduces the classical formula for the unit-propagator four-dimensional ladder integrals \cite{Usyukina:1992jd,Usyukina:1993ch} 
\begin{equation}
    \phi_{4;\gamma=1}^{(L)}(z,\zb)=\frac{1}{z-\zb}\sum_{n=0}^L\frac{(-1)^n(2L-n)!}{L!(L-n)! n!}\log(z\zb)^n(\Li_{2L-n}(z)-\Li_{2L-n}(\zb))\,,
\end{equation}
when acting with the dimensional shift operator.

Note that the dimensional recursion does not only allow us to rederive the all-loop formula for the polylogarithmic (unit-propagator power) ladder integrals in four dimensions but also gives the Fourier--Mellin representations as well as all-loop formulas for an infinite number of polylogarithmic ladder integral families with propagator powers $D/2-1$ on the horizontal and $1$ on the vertical for any even dimension $D$. For example the six-dimensional ladder integrals with propagator weights 2 on the horizontal and 1 on the vertical are given by
\begin{align}
    \phi_{6;\gamma=1}^{(L)}(z,\zb)&=\frac{1}{(z-\zb)^2}\left[\sum_{n=0}^L\frac{(-1)^n(2L-n)!}{L!(L-n)! n!}\log(z\zb)^n(\Li_{2L-n-1}(z)+\Li_{2L-n-1}(\zb)) \right. \nonumber\\
    &\left.\qquad\qquad -\frac{z+\zb}{z-\zb}\sum_{n=0}^L\frac{(-1)^n(2L-n)!}{L!(L-n)! n!}\log(z\zb)^n(\Li_{2L-n}(z)-\Li_{2L-n}(\zb))\right].
\end{align}
\section{Loop Recursion: Basso--Dixon Integrals and Toda Equations}
\label{sec:loopRecursion}

In the previous section we studied a dimensional recursion for conformal ladder integrals, induced by the operator $\dimOp$ which is proportional to the angular derivative in the two-dimensional plane parametrizing the kinematics. In this section we will now study the action of the radial derivative, corresponding to the operator $\loopOp$ already defined in the introduction, see \eqref{eq:dimAndLoopOps}. In \Secref{subsec:loopRecursionLadders} we will see that this operator induces a loop recursion on the (rescaled) four-dimensional ladder integrals with unit propagator powers. To study the action on more general four-point fishnet integrals, we will first study Basso--Dixon integrals in two dimensions in \Secref{subsec:loopRecursionBassoDixon2D} and connect the 2d Basso--Dixon formula to a two-dimensional Toda molecule equation. We will then apply the same ideas to the four-dimensional unit-propagator power Basso--Dixon integrals in \Secref{subsec:loopRecursionBassoDixon4D}, and show that they also satisfy a recursive equation, very similar to the one-dimensional Toda molecule equation. These observations hint at a connection between Basso--Dixon integrals and tau-functions, which are important objects in classical integrability. 
Before coming back to Feynman integrals, let us thus briefly review some of these integrability concepts.

\subsection{Review: Toda, Hirota and Tau-Functions}
\label{subsec:tauFcts}

The concept of tau-functions originated in the study of \textit{solitons} \cite{Zabusky:1965zz}, stable solutions of non-linear wave equations, localized in space. The first instance of a wave equation giving rise to soliton solutions, also called \textit{soliton equation}, was the Korteweg--de Vries (KdV) equation \cite{Korteweg:1895lrm}, but many more have been found including the important Kadomtsev--Petviashvili (KP) equation \cite{KPEquation} or the Toda equation \cite{todaEquation}.

Generally finding soliton solutions is a difficult task. However following the seminal work \cite{Gardner:1967wc}, introducing the \textit{inverse-scattering method}, more methods have been developed that have been found applicable across various soliton equations and remarkable unifying structures have been identified, see \cite{1791585} for a textbook introduction. In particular it has been realized that the reason for the existence of these exact soliton solutions can be traced back to a generalization of classical Liouville integrability to field theories, i.e.~systems with continuous degrees of freedom, and has led to a renaissance of the study of integrability \cite{Babelon:2003qtg}. 

One very important method for finding soliton solutions is Hirota's \textit{direct method} \cite{hirotaMethod1,hirotaMethod2}, see \cite{hirotaBook} for a pedagogical introduction. The key step of this method is to bring a soliton equation for a field $\tau(x_1,\dots,x_n)$ into the bilinear or Hirota form
\begin{equation}
    P(D_1,\dots,D_n)\tau\cdot\tau=0 \,.
\end{equation}
Here $P(D_1,\dots, D_n)$ is a function, usually a polynomial or else understood as a power series, in so-called Hirota derivatives $D_j$ with respect to the variables $x_1,\dots,x_n$. The action of each term in the polynomial or power series is defined by (for $n=2$, with obvious generalization)
\begin{equation}
    D_1^{n_1}D_2^{n_2} f\cdot g=\left.\frac{\partial^{n_1}}{\partial y_1^{n_1}}\frac{\partial^{n_2}}{\partial y_2^{n_2}}f(x_1+y_1,x_2+y_2)g(x_1-y_1,x_2-y_2)\right|_{y_1=0,y_2=0} \,.
    \label{eq:HirotaDerivative}
\end{equation}
Note that apart from continuous variables $x_j$, the above field $\tau$ may also depend on discrete lattice indices.

\paragraph{Toda Equation on the Infinite Lattice.}

An application of the above framework is provided by the traditional Toda equation \cite{todaEquation}
\begin{equation}
    \hat{\tau}_n\ddot{\hat{\tau}}_n-(\dot{\hat{\tau}}_n)^2-\hat{\tau}_{n-1}\hat{\tau}_{n+1}+\hat{\tau}_n^2=0,
    \label{eq:classToda}
\end{equation}
with dots representing time derivatives and the hat introduced to distinguish $\hat\tau$ from its rescaled version $\tau$ below. This equation corresponds to the equation of motion for an infinite one-dimensional lattice of oscillators interacting via nonlinear springs with potential
\begin{equation}
    V(r)=\frac{a}{b}\left( e^{-br}-1\right)+ar,
\end{equation}
and some parameters $a,b$. The elongation $r_n$ of the $n$-th spring  relates to the variables in the Toda equation via the equation
\begin{equation}
    \exp({r_n})-1=\frac{\dd^2}{\dd t^2}\log(\hat{\tau}_n).
\end{equation}
The above Toda equation admits the Hirota form
\begin{equation}
    \left[D_t^2 -2(\exp(D_n)-1)\right]\hat{\tau}\cdot\hat{\tau}=0 \,,
\end{equation}
where $\exp(D_n)$ acts on the discrete index $n$ labelling the springs as follows:
\begin{equation}
    \exp(D_n)\hat{\tau}\cdot\hat{\tau}=\hat{\tau}_{n-1}\hat{\tau}_{n+1} \,.
\end{equation}
This represents the discrete version of the following exponential identity which can alternatively be taken as definition of the Hirota derivative \cite{hirotaBook}:
\begin{equation}
    \exp(a D_z)f(z)\cdot g(z)=f(z+a)g(z-a) \,.
\end{equation}
Here $a$ denotes some parameter.

With an equation in Hirota form at hand it is actually relatively simple to find soliton solutions, as explained in great detail in \cite{hirotaBook}, which historically explains the plethora of new explicit soliton solutions that were found for various field equations, see e.g.~\cite{Hirota:1971zz,hirotamKdV,hirotaSineGordon}. Based on this new data it was soon realized that many soliton solutions can be written in the form of determinants, see e.g.~\cite{satsumaWronskian, Freeman:1983ih,Nimmo:1983ii}.

It was later understood \cite{satoGrassmann,Segal:1985aga} that these soliton solutions are examples of so-called \textit{tau-functions} \cite{tauFunctionsBook}, see \cite{satoReview} for an elementary account. Tau-functions are abstract mathematical objects that obey bilinear Hirota equations and are usually associated with a point in an infinite-dimensional Grassmannian manifold, the Segal--Wilson Grassmannian. Generically  they can be expressed as (infinite or finite-dimensional) determinants and it turns out that the bilinear Hirota equations they satisfy typically correspond to the Pl\"ucker relations on the infinite-dimensional Grassmannian.

Soliton solutions are not the only examples of tau-functions but this abstract mathematical framework unifies various concepts appearing in physics and mathematics, which allow for a determinant representation. Other examples include rational solutions to integrable field equations \cite{rationalTau}, certain theta-functions on Riemann surfaces \cite{thetaTau}, the Kontsevich-Witten generating function for intersection indices on moduli spaces \cite{Kontsevich:1992ti,Witten:1990hr} or matrix model integrals \cite{Kharchev:1991gd}.


\paragraph{Toda Molecule Equation on the Finite Lattice.}

In the subsequent sections we will show that four-point fishnet integrals satisfy Toda-(like) equations.
Here also open boundary conditions play a role and it turns out to be useful to rescale the above function $\hat \tau$ according to
\begin{equation}
   \tau_n=e^{a+bt+t^2/2}\hat{\tau}_n,
\end{equation}
where $a,b$ are arbitrary constants.
Hence, the above Toda equation \eqref{eq:classToda} turns into the one-dimensional Toda molecule equation \cite{Farwell:1982du,hirotaTodaMolecule}
\begin{equation}
    \tau_n\ddot{\tau}_n-(\dot{\tau}_n)^2-\tau_{n-1}\tau_{n+1}=0 \,,
    \label{eq:TodaMolecule1d}
\end{equation}
where $n$ lies in some interval, which corresponds to a finite one-dimensional lattice with one time dimension.  
Here the term \emph{molecule} refers to the presence of open boundaries.

This equation admits the Hirota form
\begin{equation}
    \left[ D_t^2-2\exp(D_n)\right]\tau\cdot\tau=0.
\end{equation}

As discussed in \Secref{subsec:loopRecursionBassoDixon4D}, the four-dimensional fishnet integrals with unit propagator powers (corresponding to $\gamma=1$) satisfy an equation similar to the above equation \eqref{eq:TodaMolecule1d}, but with an inhomogeneity on the right hand side and modified coefficients.


\paragraph{Toda Molecule Equation on the Semi-Infinite Lattice.}
Explicitly, the Basso--Dixon integrals in two dimensions, with generic deformation parameter $\gamma$, satisfy a two-dimensional (semi-infinite) Toda molecule equation \cite{toda2DSolution,Popowicz:1983pi,hirotaTodaMolecule}
\begin{equation}
    \tau_n\partial_s\partial_t\tau_n-\partial_s\tau_n\partial_t\tau_n-\tau_{n-1}\tau_{n+1}=0 \,, \qquad n\geq 0 \,,
\end{equation}
as outlined in \Secref{subsec:loopRecursionBassoDixon2D}. In this case we have a semi-infinite lattice in one spatial (discrete) but two temporal (continuous) dimensions. The above equation admits the Hirota form
\begin{equation}
    \left[ D_sD_t-2\exp(D_n) \right]\tau\cdot\tau=0.
\end{equation}

These statements which are detailed below suggest to interpret fishnet integrals in terms of the above tau-functions \cite{tauFunctionsBook}. Note that Toda equations have also shown up  \cite{Belitsky:2020qir,Olivucci:2021pss} in the context of the octagon form factor in $\mathcal{N}=4$ SYM theory \cite{Coronado:2018ypq},%
\footnote{See also \cite{Alexandrov:2011aa,Furukawa:2019piy,Kanning:2018moy,Beccaria:2022kxy} for other related occurrences of classical integrable structures in QFT amplitudes/correlation functions} 
as we will briefly review in the next subsection.

\subsection{Loop Recursions for Ladder Integrals}
\label{subsec:loopRecursionLadders}

In order to make contact between the above concepts of integrability and Basso--Dixon integrals, let us first focus on the building blocks of the 4d Basso--Dixon formula, namely on ladder integrals with unit propagator powers in four dimensions. They are given by the standard formula \cite{Usyukina:1992jd,Usyukina:1993ch}
\begin{equation}
    \phi_{4;\bm{1}}^{(L)}(z,\zb)=\frac{1}{z-\zb}L_L(z,\zb),
\end{equation}
with
\begin{equation}
    L_L(z\zb)=\sum_{n=0}^L\frac{(-1)^n(2L-n)!}{L!(L-n)! n!}\log(z\zb)^n(\Li_{2L-n}(z)-\Li_{2L-n}(\zb)).
\end{equation}
As argued in \Secref{subsec:SOVRep}, a natural representation to understand the action of the differential operators $\dimOp$ and $\loopOp$ is the Fourier--Mellin representation, i.e.~the decomposition into the respective eigenfunctions. The ladder functions $L_L(z,\zb)$ have the simple Fourier--Mellin representation\cite{Fleury:2016ykk}
\begin{equation}
   L_L(z\zb)=\sum_{n=-\infty}^\infty\int_{-\infty}^\infty\frac{\mathrm{d}\nu}{2\pi}\frac{n}{\left(\nu^2+\frac{n^2}{4} \right)^{L+1}}(z\zb)^{i\nu}\left(\frac{z}{\zb} \right)^{n/2}.
\end{equation}
Acting with the radial (logarithmic) derivative
\begin{equation}
    r\frac{\partial}{\partial r}=z\frac{\partial}{\partial z}+\zb\frac{\partial}{\partial\zb} \,,
\end{equation}
and using integration by parts we now find that (c.f.~\cite{Petkou:2021zhg,Karydas:2023ufs})
\begin{align}
    \left(z\frac{\partial}{\partial z}+\zb\frac{\partial}{\partial\zb}\right)L_L(z,\zb)&=\sum_{n=-\infty}^\infty\int_{-\infty}^{\infty}\frac{\dd\nu}{2\pi}\frac{2i\nu n}{\left(\nu^2+\frac{n^2}{4} \right)^{L+1}}(z\zb)^{i\nu}\left( \frac{z}{\zb}\right)^{n/2}\nonumber\\
    &=-\frac{1}{L}\sum_{n=-\infty}^\infty\int_{-\infty}^{\infty}\frac{\dd\nu}{2\pi}n\frac{\partial}{\partial \nu}\left(\frac{i}{\left(\nu^2+\frac{n^2}{4} \right)^{L}}\right)(z\zb)^{i\nu}\left( \frac{z}{\zb}\right)^{n/2} \nonumber\\
    &=-\frac{1}{L}\log(z\zb)\sum_{n=-\infty}^\infty\int_{-\infty}^{\infty}\frac{\dd\nu}{2\pi}\frac{n}{\left(\nu^2+\frac{n^2}{4} \right)^{L}}(z\zb)^{i\nu}\left( \frac{z}{\zb}\right)^{n/2} \nonumber\\
    &=-\frac{1}{L}\log(z\zb)L_{L-1}(z,\zb).
\end{align}
This motivates the definition of operator which was already mentioned in the introduction, cf.\ \cite{Karydas:2023ufs}:
\begin{equation}
    \loopOp=-\frac{1}{\log(z\zb)}\left(z\frac{\partial}{\partial z}+\zb\frac{\partial}{\partial\zb}\right) \,.
\end{equation}
This operator acts as a loop shift operator on the ladder integrals (more precisely the ladder functions). The same argument goes through for the regularized two-dimensional ladder integrals introduced in \Secref{subsec:SOVRep} as well as for the ladder integrals in every other even dimension which are reached by these two through the dimensional recursion.

Note that this recursion has already been observed in the two- and four-dimensional cases in \cite{Petkou:2021zhg,Karydas:2023ufs}. The Fourier--Mellin representation allows us to extend it to ladder integrals in all even dimensions for a special choice of propagator powers (the $\gamma\rightarrow 1$ limit of the one-parameter deformation \eqref{eq:propPowersOneParamDeformation}).

Recall that in four dimensions there is also another recursion in the loop order that is generated by the Laplacian $z\zb\partial_z\partial_{\zb}$ \cite{Drummond:2006rz,Drummond:2010cz} and which simply follows from the fact that the Feynman propagator is the Green's function for the four-dimensional Laplacian. The difference (and as it will turn out big upside) of the recursion presented here is that it is induced by a first order differential operator in contrast to the second order Laplacian.

Above we have restricted to four-dimensional ladder integrals with unit propagator powers (or ladder integrals related by dimensional recursion). This was due to the simple rational Fourier--Mellin representation, which is in general some ratio of $\Gamma$-functions for the parameter $\gamma$ generic (see eqs.\ \eqref{eq:2dLaddersFourierMellin,eq:4dLaddersFourierMellin}) or simply not known for more general choices of propagator weights. We were not able to find a similar loop recursion operator for other choices of $\gamma$ or $\gamma$ generic.

\subsection{Loop Recursions for 2D Basso--Dixon Integrals}
\label{subsec:loopRecursionBassoDixon2D}
In the previous subsection we have proven that there is a first order differential operator which reduces the loop order of ladder integrals with $\gamma=1$ by one. It is natural to wonder if there is a similar story for more complex four-point fishnet integrals or even general conformal integrals. In the next subsection we will show that indeed the action of this operator leads to a recursive equation for general four-dimensional four-point fishnet integrals with unit propagator powers. We will see that this equation can be identified with an inhomogeneous version of the Toda molecule equation \cite{Farwell:1982du,hirotaTodaMolecule}.

To see this we will, in this subsection, first take a step back and study the two-dimensional deformed Basso--Dixon integrals, where an exact map to a known Toda equation can be established. The techniques used here will then prove useful in the four-dimensional case.

Let us study the family of four-point fishnet integrals, see \Figref{fig:fourpointfishnetgamma},
with propagator powers differing for horizontal and vertical propagators
\begin{equation}
    a_{\mathrm{hor}}=\frac{D}{2}-\gamma,\qquad a_{\mathrm{vert}}=\gamma,
\end{equation}
with $\gamma\in(0,D/2)$ here with $D=2$. This is the generalization of the one-parameter deformation \eqref{eq:propPowersOneParamDeformation} for ladder integrals to general fishnet integrals.

As before we split off a kinematic factor carrying the conformal weights:
\begin{equation}
    I_{D=2;\gamma}^{(M,N)}=V_{D=2;\gamma}^{(M,N)}\phi_{D=2;\gamma}^{(M,N)},\qquad V_{D=2;\gamma}^{(M,N)}=\frac{\left(x_{14}^2\right)^{(\gamma-1)M}\left(x_{23}^2\right)^{(\gamma-1)M}}{\left(x_{24}^2\right)^{(\gamma-1)M+\gamma N}}.
\end{equation}
The prefactor is chosen such that the conformal function reduces to the ladder functions from before for $M=1$:
\begin{equation}
    \phi_{2;\gamma}^{(1,N)}(z,\zb)=\phi_{2;\gamma}^{(N)}.
\end{equation}
In \cite{Derkachov:2018rot} a two-dimensional version of the Basso--Dixon formula \cite{Basso:2017jwq}, expressing all fishnet integrals in terms of a determinant of (derivatives of) ladder integrals was given:
\begin{equation}
   \Phi_{M,N}:= \phi_{2;\gamma}^{(M,N)}(z,\zb)=\det_{1\leq i,j\leq M}\left( \theta^{i-1}\thetab^{j-1}\phi_{2;\gamma}^{(M+N-1)}(z,\zb)\right).
    \label{eq:2dBassoDixon}
\end{equation}
Here we employ the Euler operators $\theta=z\partial_z,\thetab=\zb\partial_{\zb}$ and we introduce the abbreviation $\Phi_{M,N}$ for the determinant to avoid clutter. This determinant formula is valid for $M\leq N$; if $M>N$ then the integral can be related to the integral with $M,N$ interchanged by permuting the external points.

\begin{figure}
    \centering
    \includegraphics[]{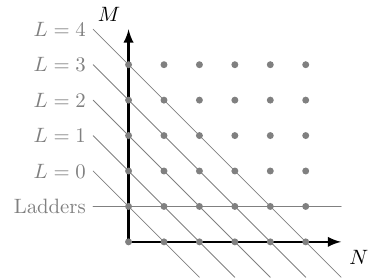}
    \caption{Basso--Dixon determinants $\Phi_{M,N}$ for fixed $L=M+N-1$ are related by the Toda equation.}
    \label{fig:TodaLattice}
\end{figure}

The form of the above determinant is referred to as a \emph{bi-directional Wronskian} and is known \cite{todaMolecule2DWronskian} to satisfy a recursive equation
\begin{equation}
    \Phi_{M,N}\theta\thetab \Phi_{M,N}-\theta\Phi_{M,N}\thetab\Phi_{M,N}-\Phi_{M+1,N-1} \Phi_{M-1,N+1}=0,
    \label{eq:toda2d}
\end{equation}
with $\Phi_{0,N}=1$. If we fix $M+N=L+1$ as well as the boundary condition 
\begin{equation}
\Phi_{0,L+1}=1,
\qquad\Phi_{1,L}=\phi_{2;\gamma}^{(L)},
\end{equation}
we can recursively compute the infinite tower of functions $\Phi_{M,L+1-M}$.%
\footnote{However only a finite number of these, namely those with $M\leq (L-1)/2$ actually correspond to Feynman integrals.} One can visualize this by foliating $(M,N)$ space into diagonal one-dimensional lattices each with a fixed $L=M+N-1$, see \Figref{fig:TodaLattice}. Then the above equation generates an entire one-dimensional lattice from the knowledge of the respective ladder.

The above recursive equation \eqref{eq:toda2d} is known as the 2d semi-infinite Toda molecule equation \cite{toda2DSolution,Popowicz:1983pi,hirotaTodaMolecule}, describing a semi-infinite lattice (molecule) with two time directions, here given by $s=\log(z),t=\log(\zb)$, cf.\ \Secref{subsec:tauFcts}.

This Toda equation for the determinants $\Phi_{M,N}$ is of course just a repackaging of the information that is already contained in the two-dimensional Basso--Dixon formula, but this representation makes a connection to classical integrability apparent and puts the fishnet integrals into the context of integrable hierarchies and tau-functions on which there is a vast literature, see e.g.\ \cite{1791585,hirotaBook,tauFunctionsBook}.

It will be instructive for the extension to the undeformed four-dimensional case to recall how the Toda equation \eqref{eq:toda2d} can be derived from the two-dimensional Basso--Dixon formula \eqref{eq:2dBassoDixon}  \cite{toda2DSolution,hirotaBook}. To this end let us fix $M+N$ and define the matrix (or rather its determinant)
\begin{equation}
    \mathsf{det}_{M,N}=\det_{1\leq i,j\leq M+1}\left( \theta^{i-1}\thetab^{j-1}\phi^{(M+N-1)}_{\gamma;2}(z,\zb)\right)=\Phi_{M+1,N-1}.
\end{equation}
One can now easily check that the constituents of the Toda equation are simply minors of the above matrix. Explicitly we have
\begin{equation}
\begin{split}
    &\Phi_{M,N}=\mathsf{det}_{M,N}\begin{bmatrix} M+1 \\ M+1 \end{bmatrix},\quad 
    \Phi_{M-1,N+1}=\mathsf{det}_{M,N}\begin{bmatrix} M & M+1 \\ M & M+1 \end{bmatrix} \,, \\
    &\theta \Phi_{M,N}=\mathsf{det}_{M,N}\begin{bmatrix} M \\ M+1 \end{bmatrix}, \quad \thetab \Phi_{M,N}=\mathsf{det}_{M,N}\begin{bmatrix} M+1 \\ M \end{bmatrix}\,,  \\
    &\theta\thetab\Phi_{M,N} = \mathsf{det}_{M,N}\begin{bmatrix} M \\ M \end{bmatrix},
\end{split}
\end{equation}
where the first (second) line in the square brackets refers to the rows (columns) eliminated before taking the determinant, i.e.~specifies the minor.

The Toda equation is then simply a consequence of an identity among the minors of an arbitrary matrix, sometimes referred to as the (Desnanot-)Jacobi identity for determinants~\cite{jacobiIdentity}
\begin{equation}
    A\begin{bmatrix} i \\ i \end{bmatrix} A\begin{bmatrix} j \\ j \end{bmatrix}
    - A\begin{bmatrix} i \\ j \end{bmatrix} A\begin{bmatrix} j \\ i \end{bmatrix}
    - A\begin{bmatrix} i & j \\ i & j \end{bmatrix} A = 0,
\end{equation}
where $A$ refers to the determinant of an arbitrary $n\times n$ matrix and $1\leq i,j \leq n$.

\subsection{Loop Recursions for 4D Basso--Dixon Integrals}
\label{subsec:loopRecursionBassoDixon4D}

We have seen that the two-dimensional four-point fishnet integrals in the one-parameter deformation satisfy a 2d Toda molecule equation reformulating the two-dimensional Basso--Dixon formula \cite{Derkachov:2018rot}. Since there is also a Basso--Dixon formula for 4d four-point fishnet integrals with unit propagator powers \cite{Basso:2017jwq}, corresponding to the $\gamma=1$ limit of the one-parameter deformation, it is natural to wonder if this can also be reformulated as a Toda equation. We will now see that this is indeed (almost) the case. Using the differential operator $\loopOp$, which we found to act as a loop-shift operator on the ladder integrals, we will derive a recursive equation for the fishnet integrals, which corresponds to a variant of the one-dimensional Toda molecule equation that contains an inhomogeneous term and some non-trivial coefficients.

First let us state our conventions. We will, as before, define the conformal functions associated to the fishnet integrals by stripping off a kinematic factor carrying the conformal weights
\begin{equation}
    I_{D=4;\gamma=1}^{(M,N)}=V_{D=4;\gamma=1}^{(M,N)}\phi_{D=4;\gamma=1}^{(M,N)},\quad V_{D=4;\gamma=1}^{(M,N)}=\frac{\left(x_{24}^2 \right)^{M-N}}{\left(x_{14}^2 \right)^{M}\left(x_{23}^2 \right)^{M}},
\end{equation}
chosen such that
\begin{equation}
    \phi_{4;1}^{(1,N)}=\phi_{4;1}^{(N)}.
\end{equation}
As conjectured by Basso and Dixon \cite{Basso:2017jwq} and later proven \cite{Basso:2021omx}, the conformal functions are given by determinants of ladder functions
\begin{equation}
    \phi_{4;1}^{(M,N)}=\frac{1}{(z-\zb)^M}\frac{1}{\prod_{k=0}^{2M-1}(N-M+k)!}\det_{1\leq i,j\leq M}\left( f_{N-M+i+j-1}(z,\zb)\right).
\end{equation}
Here we made use of the rescaled ladder functions of \cite{Coronado:2018ypq,Coronado:2018cxj} given by
\begin{equation}
\label{eq:rescaledLadder}
    f_L(z,\zb)=L!(L-1)!L_L(z,\zb)=\sum_{j=L}^{2L}\frac{(-1)^j (L-1)!j!}{(j-L)!(2L-j)!}\left( \log(z\zb)\right)^{2L-j}\left( \mathrm{Li}_j(z)-\mathrm{Li}_j(\zb)\right).
\end{equation}

The key step to reformulate the Basso--Dixon formula is now to rewrite the determinant using the loop recursion which, for the rescaled ladder functions, takes the form
\begin{equation}
    \loopOp f_L(z,\zb)\equiv -\frac{1}{\log(z\zb)}(z\partial_z+\zb\partial_{\zb})f_L(z,\zb)=(L-1)f_{L-1}(z,\zb),\qquad L>1.
\end{equation}
We can thus express
\begin{equation}
   \phi_{4;1}^{(M,N)}=\frac{1}{(z-\zb)^M}\frac{1}{\left[(M+N-2)!\right]^M\prod_{k=0}^{2M-1}(N-M+k)!}\Psi_{M,N},
\end{equation}
where we abbreviate
\begin{equation}
\label{eq:psiBassoDixon}
    \Psi_{M,N}=\det_{1\leq i,j\leq M}\left( c_{i+j}\loopOp^{i+j-2}f_{N+M-1}(z,\zb)\right) \,.
\end{equation}
Here the coefficients are given by
\begin{equation}
    c_k=(M+N-k)! \,.
\end{equation}
Note that these depend on $M+N$, but since this sum will always be fixed in the following we will suppress this dependence to avoid clutter.

Apart from the coefficients $c_k$ the above equation looks very much like a one-variable version of the Basso--Dixon formula for two-dimensional fishnet integrals. We can hence try to take similar steps as before and derive a recursion relation for the four-dimensional unit-propagator power fishnet integrals. Due to the coefficients this will not exactly map to the known form of the one-dimensional Toda molecule equation but be very similar. Motivated by the 2d discussion let us fix $K=M+N$ and define the determinant
\begin{equation}
    \mathsf{det}_{M,N}=\det_{1\leq i,j\leq M+1}\left(c_{i+j}\loopOp^{i+j-2}f_{N+M-1}(z,\zb)\right)=\Psi_{M+1,N-1}.
\end{equation}
We then find analogously to before
\begin{equation}
    \Psi_{M,N}=\mathsf{det}_{M,N}\begin{bmatrix} M+1 \\ M+1  \end{bmatrix},\qquad
    \Psi_{M-1,N+1}=\mathsf{det}_{M,N}\begin{bmatrix} M & M+1 \\ M & M+1  \end{bmatrix}.
\end{equation}
For the derivatives, however, the presence of the coefficients $c_k$ makes a difference and we have
\begin{align}
    \loopOp\Psi_{M,N}&= \frac{c_{2M}}{c_{2M+1}}\mathsf{det}_{M,N}\begin{bmatrix} M+1 \\ M \end{bmatrix},\quad N>M \\
    \loopOp^2\Psi_{M,N}&=\frac{c_{2M}}{c_{2M+2}}\left(\mathsf{det}_{M,N}\begin{bmatrix} M \\ M \end{bmatrix} +\mathcal{M}_{M,N}\right),\quad N>M+1 \,.
    \label{eq:extraterm}
\end{align}
We see that the presence of the coefficients in the Basso--Dixon determinant not only introduces coefficients into the relations but also leads to the extra term on the right hand side of \eqref{eq:extraterm}. 
While we are lacking a good interpretation of this inhomogeneuos term in the context of a standard Toda equation, we were able to experimentally find a formula for it in terms of a minor of ladder integrals (verified up to $M=8,N=20$)
\begin{equation}
    \mathcal{M}_{M,N}=\frac{2c_{2M+2}c_2^M}{c_{2M+1}}\det_{1\leq i,j\leq M+1}\left( f_{i+j+N-M-2}\right)\begin{bmatrix} 3 \\ M+1 \end{bmatrix} \,,
\end{equation}
valid for $N>M+1$. Note that we need to set $\mathcal{M}_{1,N}=0$ for $N>2$. For $M=2$ we can actually identify the inhomogeneity as another Basso--Dixon integral
\begin{equation}
    \mathcal{M}_{2,N}=\frac{2N^2}{N-3}\Psi_{2,N-1},\quad N=3 \,.
\end{equation}

Using the above results and the Jacobi identity for determinants we can now, analogously to before, deduce a recursive equation for the four-dimensional undeformed fishnet integrals
\begin{align}
    \frac{c_{2M+2}}{c_{2M}}\Psi_{M,N}\loopOp^2\Psi_{M,N}-\left( \frac{c_{2M+1}}{c_{2M}}\right)^2\left(\loopOp\, \Psi_{M,N} \right)^2-\Psi_{M-1,N+1}\Psi_{M+1,N-1} 
    =\Psi_{M,N}\mathcal{M}_{M,N} ,
    \label{eq:Toda4d}
\end{align}
for $N>M+1$. Note that the latter restriction precisely guarantees that the $\Psi_{M,N}$ entering the equation (in particular $\Psi_{M+1,N-1}$) have their second index larger than their first, ensuring that they are captured by the Basso--Dixon formula \eqref{eq:psiBassoDixon}.

The recursive equation for $\Psi_{M,N}$ looks very similar to the one-dimensional Toda molecule equation, cf.\ \eqref{eq:TodaMolecule1d} and \cite{Farwell:1982du,hirotaTodaMolecule}, however with non-trivial coefficients and a \enquote{source term} on the right hand side. Hence, the connection to standard classically integrable models is less clear in the 4d case. Nonetheless, exactly as for the two-dimensional fishnet integrals, this is a recursive equation allowing one to find all four-dimensional fishnet integrals with fixed $N+M=L+1$ from the boundary condition $\Psi_{0,N}=1$ and the knowledge of the corresponding ladder integral~$\Psi_{1,L}$.

Given the findings of this section, namely that two- and four-dimensional Basso--Dixon integrals satisfy Toda(-like) equations, 
it is natural to understand these integrals in the context of tau-functions. As the Toda equations are a classic example of integrable field equations \cite{todaIntegrability,Takasaki:2018wsv}, their connection to tau-functions is contained in the literature.
In particular, the two-dimensional Toda molecule equation has been studied in this context in \cite{toda2DMoleculeAndTau}, see also \cite{toda2DAndTau}. There it was found that this equation admits determinant solutions that can be interpreted as tau-functions. These take a \textit{double Wronskian} form
\begin{equation}
    \tau_n(x,y)=\det\begin{pmatrix}
        \phi_1 & \partial_x \phi_1 & \dots & \partial_x^{n-1}\phi_1 & \psi_1 & \partial_y\psi_1 & \dots & \partial_y^{M-n-1}\psi_1 \\
        \phi_2 & \partial_x \phi_2 & \dots & \partial_x^{n-1}\phi_2 & \psi_2 & \partial_y\psi_2 & \dots & \partial_y^{M-n-1}\psi_2 \\
        \vdots & \vdots &  & \vdots & \vdots & \vdots &  & \vdots \\
        \phi_M & \partial_x \phi_M & \dots & \partial_x^{n-1}\phi_M & \psi_M & \partial_y\psi_M & \dots & \partial_y^{M-n-1}\psi_M 
    \end{pmatrix}\,,
\end{equation}
for arbitrary functions $\phi_i(x),\psi_i(y)$, $i=1,\dots,M$. The Toda molecule equation can then be interpreted as a Pl\"ucker relation, see \cite{toda2DMoleculeAndTau} for details. The fishnet integrals however take a \textit{bi-directional Wronskian} form. These two can be related as discussed in \cite{toda2DMoleculeAndTau}, though we are not aware of an explicit discussion of Pl\"ucker relations for the latter.

The one-dimensional Toda molecule equation, which looks similar to the equation obeyed by the 4d Basso--Dixon determinants, has been shown to admit determinant solutions (of Hankel type) \cite{toda1DHankel,toda1DHankel2}; again we are not aware of  an explicit discussion of Pl\"ucker relations in this context.

It is an intriguing question how the Basso--Dixon formulas are embedded into the framework of tau-functions, and which other quantum field theory observables might fit into this picture.%
\footnote{See also \cite{Beccaria:2022ypy,Bajnok:2024epf} in this context.}
Let us briefly review an interesting connection with regard to the latter question.


\paragraph{Basso--Dixon Integrals and the Octagon.}

The form of the above variant \eqref{eq:Toda4d} of the Toda equation for 4d Basso--Dixon integrals is somewhat mysterious. Notably, those integrals can be understood as special limits of the so-called octagon form factor, which also in another limit was found to obey a Toda equation. Let us therefore briefly review this connection in the following.

The octagon form factor $\mathbb{O}_l(z,\zb,\alpha,\bar{\alpha})$ in $\mathcal{N}=4$ super Yang-Mills theory \cite{Coronado:2018ypq} is an important building block for four-point correlation functions of single-trace half BPS operators $\mathcal{O}_i=\Tr\left( (y_i\cdot \Phi(x_i))^K\right)$, with certain choices of $\mathrm{SO}(6)$ polarizations $y_i$ and large $R$-charge $K\gg 1$, see also \cite{Bargheer:2019kxb}. Here $\Phi$ denotes the R-symmetry $\mathrm{SO}(6)$ vector of scalar fields of the $\mathcal{N}=4$ supermultiplet. The octagon form factor depends on the conformal cross ratios formed out of the four insertion points $x_1,\dots,x_4$, 
defined as in \eqref{eq:conformalCrossRatios}, as well as R-symmetry cross ratios formed out of the four polarization vectors
\begin{equation}
    \alpha \bar{\alpha}=\frac{(y_1\cdot y_2)(y_3\cdot y_4)}{(y_1\cdot y_4)(y_2\cdot y_3)},
    \qquad (1-\alpha)(1-\bar{\alpha})=\frac{(y_1\cdot y_3)(y_2\cdot y_4)}{(y_1\cdot y_4)(y_2\cdot y_3)}\, .
\end{equation}

At every order in perturbation theory the octagon form factor can be written in terms of determinants of the ladder functions $f_L(z,\zb)$, see eq.~\eqref{eq:rescaledLadder}. This was first observed by explicit perturbative computations in \cite{Coronado:2018ypq,Coronado:2018cxj} and then explicitly proven in \cite{Kostov:2019stn,Kostov:2019auq} from a finite-coupling determinant representation of the octagon, derived from hexagonalization \cite{Basso:2015zoa,Fleury:2016ykk}. Explicitly, the weak-coupling expansion of the octagon form factor reads \cite{Kostov:2019stn,Kostov:2019auq}
\begin{equation}
    \mathbb{O}_l=\frac{1}{2}\sum_{\sigma=\pm}\det_{0\leq i,j\leq N-1}\left( \delta_{i,j}+\mathcal{X}_\sigma e^{-\xi}\mathcal{R}_{ij}(z,\zb,g)\right)+\mathcal{O}\left(g^{4N+2l}\right) \, ,
\end{equation}
where
\begin{equation}
    \mathcal{X}_+=-\frac{(z-\alpha)(\zb-\alpha)}{\alpha},\qquad \mathcal{X}_-=-\frac{(z-\bar{\alpha})(\zb-\bar{\alpha})}{\bar{\alpha}}, \qquad \xi=-\frac{1}{2}\log(z\zb) \, ,
\end{equation}
as well as
\begin{equation}
    \mathcal{R}_{ij}(z,\zb,g)=\sum_{p=\max(i+j+l,1+j+l)}^{2N}r_{i,j}(p)f_p(z,\zb)g^{2p} \, ,
\end{equation}
with coefficients given by
\begin{equation}
    r_{i,j}(p)=\frac{(-1)^{p-l}(2p-1)!\left[2p(2i+l)(1-\delta_{i,0})-\delta_{i,0}(p-j-l)(j+p) \right]}{(i-j+p)!(j-i+p)!(p-i-j-l)!(p+i+j+l)!}.
\end{equation}
While in principle various determinants of ladder integrals show up in the perturbative expansion, the connection to fishnet integrals in particular can be made even more explicit by organising the perturbative series as follows \cite{Kostov:2019stn}
\begin{align}
    \mathbb{O}_l&=\sum_{M=0}^\infty \mathcal{X}_M g^{2M(M+l)}\left[\frac{1}{\prod_{i=1}^M(2i+l-2)!(2i+l-1)!}\det_{1\leq i,j\leq M}\left(f_{i+j+l-1} \right)+\mathcal{O}(g^2) \right] \nonumber \\
    &=\sum_{M=0}^\infty \mathcal{X}_M g^{2M(M+l)}\left[\frac{\Psi_{M,M+l}}{[(2M+l-2)!]^M\prod_{i=1}^M(2i+l-2)!(2i+l-1)!}+\mathcal{O}(g^2) \right],
\end{align}
with 
\begin{equation}
    \mathcal{X}_M=\frac{1}{2}(\mathcal{X}_+^M+\mathcal{X}_-^M)e^{-M\xi}.
\end{equation}
Since we have seen that the $\Psi_{M,N}$ (and more generally determinants of ladder functions) satisfy Toda-like equations it is a natural question if this hints at a Toda-like equation obeyed by the full octagon form factor (or some other Hirota equation, see \Secref{subsec:tauFcts}).

As a further hint into this direction, in \cite{Belitsky:2020qir} it was shown, using a different finite-coupling determinant representation \cite{Belitsky:2019fan,Belitsky:2020qrm}, that in a certain light-cone double-scaling limit the (rescaled) octagon satisfies a one-dimensional Toda equation. In this limit the coupling becomes small and the four operators in the correlation function approach the configuration of a null square. Explicitly one takes the limit
\begin{equation}
    z\rightarrow 0, \quad \zb\rightarrow \infty,\quad g\rightarrow 0, \quad \text{with}\quad \xi=-\frac{1}{2}\log(z\zb), \quad s=-g\log\left(\frac{z}{\zb}\right) \quad\text{both fixed.}
\end{equation}
In this limit the rescaled octagon
\begin{equation}
    \tau_l(s)=s^{l^2}e^{s^2/4}\,\mathbb{O}_l \, ,
\end{equation}
satisfies a one-dimensional Toda equation in the variable $t=\log(s)$
\begin{equation}
    \tau_l(t)\ddot{\tau}_l(t)-\dot{\tau}_l^2(t)-\tau_{l-1}(t)\tau_{l+1}(t)=0 \, ,
\end{equation}
with the solution
\begin{equation}
    \tau_l(s)=s^{l^2}\det_{1\leq i,j\leq l}(I_{i-j}(s)) \, ,
\end{equation}
where $I_k(s)$ denotes the modified Bessel function of the first kind.
It was further found that the octagon also satisfies a two-dimensional Toda equation in a light-cone single scaling limit, see \cite{Olivucci:2021pss} for details.

While the Toda-like equation for fishnet integrals and the Toda equations satisfied by the octagon in certain limits are certainly good hints that there might be a similar equation for the full octagon, we have not managed to make this explicit so far and we will leave this to future investigation.

\section{Double Copy Form of Four-Point Integrals}
\label{sec:doubleCopy}
In this section we will investigate another implication of the dimensional recursion found in \Secref{sec:dimRecursion}. For integrals in two dimensions it is well-known that they can be brought to a double copy form which will be reviewed in \Secref{sec:2dDoubleCopy}. 

Using the dimensional recursion we can also bring 4d integrals into a double copy form, which will be explained in \Secref{sec:4dDoubleCopy} and illustrated on concrete examples in the rest of the section.

\subsection{Review of Double Copy in Two Dimensions}
\label{sec:2dDoubleCopy}
It is well known that physics can be dramatically different and simpler in two dimensions. This usually can be traced back to the fact that in two dimensions we can change variables to the (anti-)holomorphic variables
\begin{equation}
    z=x^1+i x^2,\qquad \zb =x^1-ix^2\, ,
\end{equation}
which allows the use of the powerful tools of complex analysis. A prime example of this is two-dimensional conformal field theory \cite{DiFrancesco:1997nk}, where even the full conformal algebra splits into holomorphic and anti-holomorphic parts. In particular important quantities like conformal blocks take the form of a product of a holomorphic and an anti-holomorphic function. This immediately implies a factorization of any two-dimensional conformal correlation function as it can be expanded in terms of conformal blocks and can hence be written as an (infinite) sum of factorized terms.

For two-dimensional Feynman integrals one can make an even stronger statement, even in the absence of (dual) conformal symmetry. Again the introduction of (anti-)holomorphic variables is at the heart of the simplification. We will do so both for the external points $x_k$ and the integration variables $y_k$:
\begin{equation}
    z_k=x_k^1+i x_k^2,\qquad w_k=y_k^1+i y_k^2 \, .
\end{equation}
In these variables, both the integration measure
\begin{equation}
    \frac{1}{\pi^L}\bigwedge_{k=1}^L\left( \mathrm{d}x_k^1\wedge \mathrm{d}x_k^2\right)=\left(- \frac{1}{2\pi i}\right)^L\bigwedge_{i=1}^L\left( \mathrm{d}w_i\wedge\mathrm{d}\Bar{w}_i \right) \, ,
\end{equation}
as well as the propagator factors
\begin{equation}
    D_i=\left(\sum_j \alpha_{i,j}x_j+\sum_i\beta_{i,j}y_j \right)^2=\left|\sum_j \alpha_{i,j}z_j+\sum_i\beta_{i,j}w_j \right|^2 \,,
\end{equation}
factorize into a holomorphic and an anti-holomorphic part. Here $L$ is the loop order and $\alpha_{i,j},\beta_{i,j}\in\{-1,0,1\}$ are determined by the topology of the graph. 

This simple observation implies that the integrand of any two-dimensional massless Feynman integral factorizes and suggests that there might also be some factorization into holomorphic and anti-holomorphic parts in the result. Indeed in \cite{Duhr:2023bku} this observation is made rigorous and it is proven that every $n$-point massless two-dimensional Feynman integral can be expressed in a double-copy form
\begin{equation}
    \phi(\bm{z},\bm{\zb})=\Pi(\bm{z})^\dagger\Sigma\Pi(\bm{z}).
\end{equation}
where $\bm{z}=(z_1,\dots,z_n),\bm{\zb}=(\zb_1,\dots,\zb_n)$ and $\phi(\bm{z},\bm{\zb})$ is the conformal function obtained by stripping off a kinematic factor carrying the scaling weights of the Feynman integral. Furthermore $\Pi(\bm{z})$, the so-called period vector, is a vector of hypergeometric functions (at least for generic propagator powers) of the holomorphic variables $\bm{z}$ and $\Sigma$, the so-called intersection matrix, is a constant matrix.

Hence indeed the Feynman integral can be written as a linear combination of factorized products where the individual functions only depend on the (anti-)holomorphic variables. The intersection form is further invariant under all monodromies of the period vector, leading to $\phi(\bm{z},\bm{\zb})$ being single-valued. Explicitly, if we take some kinematic parameter around some closed contour $C$ in kinematic space, then in general the period vector will undergo some rotation
\begin{equation}
    \Pi(\bm{z})\rightarrow M_C\Pi(\bm{z}) \, .
\end{equation}
The intersection matrix then satisfies
\begin{equation}
    M_C^\dagger \Sigma M_C= \Sigma \, ,
\end{equation}
leading to the Feynman integral being invariant.

These single-valued combinations of hypergeometric functions have been systematically studied in the framework of twisted (co-)homology \cite{aomoto,yoshida}, leading to explicit expressions for single-valued versions of many well-known hypergeometric functions like the generalized hypergeometric functions ${}_{p+1} F_p$ \cite{svpFq1,svpFq2} or the 
Lauricella functions $F_D^{(r)}$ \cite{Brown:2019jng}, see \cite{Duhr:2023bku} for a nice review in the physics literature.

Note that we have already seen this double copy form for two-dimensional integrals with certain rational propagator powers in \Secref{subsec:CYlift}, where we briefly reviewed results from \cite{Duhr:2022pch,Duhr:2023eld,Duhr:2024hjf}. In the references it was shown that this double copy form naturally relates to the underlying Calabi--Yau geometry, which at the same time gives a systematic way to compute power-series representations for the period vector as well as find a simple form for the intersection matrix. The result of \cite{Duhr:2023bku} shows that, despite the Calabi--Yau techniques not being applicable for more general choices of propagator powers, this double copy form is something completely general for massless two-dimensional integrals.

Let us illustrate this on the simple example of the box integral in two dimensions\footnote{Note the unusual propagator power assignments in accord with our earlier conventions.}
\begin{equation}
    \phi_{D=2;\mathbf{a}}^{(1)}(z,\zb)=\left(V_{D=2;\mathbf{a}}^{(L)}(z,\zb)\right)^{-1}\int\frac{\mathrm{d}^2 y}{\pi}\frac{1}{(x_1-y)^{2a_1}(x_2-y)^{2a_3}(x_3-y)^{2a_2}(x_4-y)^{2a_4}}\, ,
\end{equation}
with the kinematic prefactor $V_{\mathbf{a}}$ as before, see eq.~\eqref{eq:VFactorLadder}. As seen before, the integrand manifestly factorizes when changing to complex variables, suggesting the double copy form. By some simple transformations we can actually immediately identify the integral as the single-valued version of the familiar Gauss hypergeometric function. To this end we make a conformal transformation to send $x_4\rightarrow\infty$ and after some straightforward manipulations we find
\begin{equation}
    \phi_{2;\mathbf{a}}^{(1)}(z,\zb)=-\frac{1}{2\pi i}\int_\mathbb{C}\mathrm{d}w\wedge\mathrm{d}\bar{w}\frac{1}{|1-w|^{2a_1}|1-z w|^{2a_2}|w|^{2a_3}}\, ,
\end{equation}
which on the level of the integrand already looks like a double copy of the integral representation of the Gauss hypergeometric function ${}_2 F_1$, cf.\ e.g.\ \cite{aomoto}. Indeed this is the integral representation of its single-valued version
\begin{equation}
    \phi_{2;\mathbf{a}}^{(1)}(z,\zb)=\psvFq{2}{1}{a_2,\,1-a_3}{2-a_1-a_3}{z} \, ,
\end{equation}
using the conformal constraint $\sum_{i=1}^4a_i=2$ to eliminate $a_4$. Single-valued hypergeometric functions can be written as bilinears in ordinary hypergeometric functions, see \appref{sec:svHGFs}. Explicitly we can write the single-valued Gauss hypergeometric function as
\begin{equation}
    \psvFq{2}{1}{\alpha,\,\beta}{\gamma}{z}=\Pi_{\mathrm{G}}(z)^\dagger\Sigma_{\mathrm{G}}\Pi_{\mathrm{G}}(z) \, ,
\end{equation}
with the period vector
\begin{equation}
    \Pi_{\mathrm{G}}=\begin{pmatrix}
        \pcFq{2}{1}{\alpha,\,\beta}{\gamma}{z} \\
        z^{1-\gamma}\pcFq{2}{1}{1+\alpha-\gamma,\,1+\beta-\gamma}{2-\gamma}{z}
    \end{pmatrix}\, ,
\end{equation}
and the intersection matrix
\begin{equation}
    \Sigma_{\mathrm{G}}=\begin{pmatrix}
        \frac{\sin(\pi\beta)\sin(\pi(\gamma-\beta))}{\pi\sin(\pi\beta)} & 0 \\
        0 & -\frac{\sin(\pi\alpha)\sin(\pi(\gamma-\alpha))}{\pi\sin(\pi\gamma)}
    \end{pmatrix}\, .
\end{equation}
Here the rescaled Gauss function is defined by
\begin{equation}
    \pcFq{2}{1}{\alpha,\,\beta}{\gamma}{z}= \frac{\Gamma(\beta)\Gamma(\gamma-\beta)}{\Gamma(\gamma)}\pFq{2}{1}{\alpha,\,\beta}{\gamma}{z} \, .
\end{equation}


\subsection{Double Copy in Higher Dimensions from Recursions}
\label{sec:4dDoubleCopy}
In the previous subsection we reviewed that we can generally write two-dimensional massless Feynman integrals as a double copy
\begin{equation}
    \phi_{D=2}(\bm{z},\bm{\bar{z}})=\Pi(\bm{z})^\dagger\Sigma\Pi(\bm{z}).
\end{equation}
Focussing on the case of conformal ladder integrals we can now extend this statement to higher dimensions simply by making use of the recursion from \Secref{sec:dimRecursion}, repeated here for convenience
\begin{equation}
    \phi_{D=4;\bm{a}}^{(L)}=\frac{1}{\prod_{i=1}^{L+1}(a_i-1)}\frac{1}{z-\zb}(z\partial_z-\zb \partial_{\zb})\phi_{D=2;\bm{a}-\bm{e}_{1,2,\dots,L+1}}^{(L)}.
\end{equation}
By simply plugging in the double copy form for the two-dimensional integral
\begin{equation}
    \phi_{2;\bm{a}}^{(L)}(z,\zb)=\Pi_2^{(L)}(z;\bm{a})^\dagger\Sigma_2^{(L)}(\bm{a})\Pi_2^{(L)}(z;\bm{a})   
\end{equation}
we find for the four-dimensional ladder integrals
\begin{align}
    \phi_{4;\bm{a}}^{(L)}(z,\zb)&=\frac{1}{\prod_{i=1}^{L+1}(a_i-1)}\frac{1}{z-\zb}\left[ \Pi_2^{(L)}(z;\bm{a})^\dagger\Sigma_2^{(L)}(\bm{a})z\partial_z\Pi_2^{(L)}(z;\bm{a}) \right. \nonumber\\
    &\left.\qquad\qquad\qquad\qquad\qquad  -\left(z\partial_z \Pi_2^{(L)}(z;\bm{a})\right)^\dagger\Sigma_2^{(L)}(\bm{a})\Pi_2^{(L)}(z;\bm{a})  \right] \,,
\end{align}
which we can repackage into the double copy form
\begin{equation}
    \phi_{4;\bm{a}}^{(L)}(z,\zb)=\frac{1}{\prod_{i=1}^{L+1}(a_i-1)}\frac{1}{z-\zb}\Pi_4^{(L)}(z;\bm{a})^\dagger\Sigma_4^{(L)}(\bm{a})\Pi_4^{(L)}(z;\bm{a}) \, ,
\end{equation}
with
\begin{equation}
    \Pi_4^{(L)}(z;\bm{a})=\begin{pmatrix} \Pi_2^{(L)}(z;\bm{a}-\bm{e}_{1,2,\dots,L+1}) \\ z \partial_z\Pi_2^{(L)}(z;\bm{a}-\bm{e}_{1,2,\dots,L+1}) \end{pmatrix} \, ,
\end{equation}
and
\begin{equation}
    \Sigma_4^{(L)}(\bm{a})=\begin{pmatrix} 0 && \Sigma_2^{(L)}(\bm{a}-\bm{e}_{1,2,\dots,L+1}) \\ -\Sigma_2^{(L)}(\bm{a}-\bm{e}_{1,2,\dots,L+1}) && 0 \end{pmatrix} \, .
\end{equation}
We see that the four-dimensional ladder integrals have a similar double copy structure with the period vector being composed of the two-dimensional periods and their derivatives (and shifted propagator powers). Note that in concrete examples it might be useful to employ a different basis for the period vector. Furthermore it should be clear that we can extend the above argument to any even dimension by repeatedly applying the recursion.

Since the double copy form of two-dimensional Feynman integrals is very general, the same conclusion applies to higher-point integrals with the usual caveats. For every track-like integral the recursion implies that it can be written in a double copy form in any even dimension when restricted to two-dimensional kinematics. 

In special cases we can actually extend the double copy form to more non-trivial fishnet integrals. Namely, if there is a Basso--Dixon like formula expressing fishnet integrals as determinants in (derivatives of) ladder integrals, then the double copy form of the ladder integrals implies a double copy form for the general fishnet integrals. The original Basso--Dixon formula hence implies a double copy form for all four-dimensional fishnet integrals with unit propagator powers. We will illustrate this in the example of the window integral in \Secref{subsec:windowDoubleCopy}.

There is actually a subtlety here concerning the unit propagator power limit. The period vector might actually diverge in this limit, with the poles cancelled by zeros of the intersection matrix. This leads to a sum of double copy terms which in particular can mix different orders in the expansion around unit propagator weights. This however does not change the conclusion that the unit-propagator weight integrals have a double copy form. We will illustrate this in the subsequent \Secref{subsec:boxDoubleCopy} on the example of the box integral.

\subsection{Example: Box Integral}
\label{subsec:boxDoubleCopy}
As a first example we consider the box integral. As already reviewed in the last subsection, the two-dimensional box integral has the double copy form
\begin{equation}
    \phi_{2;\bm{a}}^{(1)}(z,\zb)=\Pi_2^{(1)}(z;\bm{a})^\dagger\Sigma_2^{(1)}(\bm{a})\Pi_2^{(1)}(z;\bm{a}) \, ,
\end{equation}
with the period vector
\begin{equation}
    \Pi_2^{(1)}(z;\bm{a})=\begin{pmatrix}
        \pcFq{2}{1}{a_2,\,1-a_3}{2-a_1-a_3}{z} \\
        z^{a_1+a_3-1}\pcFq{2}{1}{a_1+a_2+a_3-1,\,a_1}{a_1+a_3}{z} 
    \end{pmatrix},
\end{equation}
and the intersection matrix
\begin{equation}
    \Sigma_2^{(1)}(\bm{a})=\begin{pmatrix}
        \frac{\fs(1-a_1)\fs(1-a_3)}{\pi\fs(2-a_{13})} & 0 \\
        0 & -\frac{\fs(a_2)\fs(2-a_{123})}{\pi\sf(2-a_{13})}
    \end{pmatrix} \, ,
\end{equation}
where we introduced the abbreviations $\mathfrak{s}(x)=\sin(\pi x)$ and $a_{i_1,\dots ,i_k}=a_{i_1}+\dots +a_{i_k}$. Applying the recursion operator we then find a double copy form for the four-dimensional box integral
\begin{equation}
    \phi_{4;\bm{a}}^{(1)}(z,\zb)=\frac{1}{(a_1-1)(a_2-1)}\frac{1}{z-\zb}\Pi_4^{(1)}(z;\bm{a})^\dagger\Sigma_4^{(1)}(\bm{a})\Pi_4^{(1)}(z;\bm{a}) \, ,
\end{equation}
with the period vector
\begin{equation}
    \Pi_4^{(1)}(z;\bm{a})=\begin{pmatrix}
        \pcFq{2}{1}{a_2-1,\,1-a_3}{3-a_1-a_3}{z} \\
        z\,\pcFq{2}{1}{a_2,\,2-a_3}{4-a_1-a_3}{z} \\
        z^{a_1+a_3-2} \pcFq{2}{1}{a_1-1,\,a_1+a_2+a_3-3}{a_1+a_3-1}{z} \\
        z^{a_1+a_3-1} \pcFq{2}{1}{a_1,\,a_1+a_2+a_3-2}{a_1+a_3}{z} 
    \end{pmatrix},
\end{equation}
and the intersection matrix
\begin{equation}
    \Sigma_4^{(1)}(\bm{a})= \begin{pmatrix}
        0 & \frac{(a_2-1)\mathfrak{s}(a_1)\mathfrak{s}(a_3)}{\pi \mathfrak{s}(a_{13}-3)} & 0 & 0 \\
        -\frac{(a_2-1)\mathfrak{s}(a_1)\mathfrak{s}(a_3)}{\pi \mathfrak{s}(a_{13}-3)} & 0  & 0 & 0 \\
        0 & 0 &  0 & \frac{(a_1-1)\mathfrak{s}(a_2)\mathfrak{s}(a_{123}-4)}{\pi \mathfrak{s}(a_{13}-3)} \\
        0 & 0 & -\frac{(a_1-1)\mathfrak{s}(a_2)\mathfrak{s}(a_{123}-4)}{\pi \mathfrak{s}(a_{13}-3)} & 0
    \end{pmatrix} \, .
\end{equation}

Let us consider the unit-propagator limit of the above formula which leads us to a double-copy formula for the Bloch-Wigner dilogarithm which is famously single-valued. If we set $a_j=1+j\epsilon,~j=1,2,3$ and expand in $\epsilon$ 
\begin{align}
    \Pi_4^{(1)}(z;\bm{a})&=\frac{1}{\epsilon}\pi_{-1}(z)+\pi_{0}(z)+\pi_1(z)\epsilon+\mathcal{O}(\epsilon^2) \,, \\
    \frac{1}{(a_1-1)(a_2-1)}\Sigma_4^{(1)}(\bm{a})&=\sigma_0+\sigma_2\epsilon^2+\mathcal{O}(\epsilon^4) \,.
\end{align}
We find that the double copy form splits up into three terms
\begin{equation}
    \phi_{4;\bm{1}}^{(1)}(z,\zb)=\frac{1}{z-\zb}\left[\pi_0(z)^\dagger\sigma_0\pi_0(z)+\pi_{-1}(z)^\dagger\sigma_0\pi_1(z)+\pi_1(z)^\dagger\sigma_0\pi_{-1}(z) \right] \,.
    \label{eq:boxDoubleCopyExp}
\end{equation}
The relevant Laurent coefficients are explicitly given by
\begin{align}
    \pi_{-1}(z)&=\left(-\frac{1}{3},0,\frac{1}{6},0\right) \,, \\
    \pi_{0}(z)&=\left(0,-\log(1-z),\frac{2}{3}\log(z),-\log(1-z)\right) \,, \\
    \pi_{1}(z)&=\left(\frac{\pi^2}{6}+2\Li_2(z), \frac{3}{2}\log(1-z)^2+4\Li_2(z),\frac{\pi^2}{3}+\frac{4}{3}\log(z)^2+\Li_2(z),\right.\\
     &\qquad\left.\frac{3}{2}\log(1-z)^2-4\log(1-z)\log(z)-4\Li_2(z) \right)\,,
\end{align}
and
\begin{equation}
    \sigma_0=\frac{1}{4}\begin{pmatrix}
        0 & -3 & 0 & 0 \\
        3 & 0 & 0 & 0 \\
        0 & 0 & 0 & -6 \\
        0 & 0 & 6 & 0
    \end{pmatrix} \,.
\end{equation}
Note that the periods in $\pi_0$ correspond to the solutions of the Yangian Ward identities \cite{Loebbert:2019vcj} for the two-dimensional conformal box integral with propagator powers $1$ on the vertical and $0$ on the horizontal.

Plugging the expansions of the periods and the intersection matrix into \eqref{eq:boxDoubleCopyExp} one easily recovers the well-known expression of the four-dimensional unit propagator power box integral in terms of the Bloch-Wigner dilogarithm
\begin{equation}
    \phi_{4;\bm{1}}^{(1)}(z,\zb)=\frac{1}{z-\zb}\left[2\Li_2(z)-2\Li_2(\zb)+\log(z\zb)\log\left(\frac{1-z}{1-\zb} \right) \right] \,.
\end{equation}
\subsection{Example: Two-loop Ladder Integral}
\label{sec:2loopLadder}
As a second example we will study the two-loop ladder integral. 
In two dimensions the double ladder has the double copy form
\begin{equation}
    \phi_{2;\bm{a}}^{(2)}(z,\zb)=\Pi_2^{(2)}(z,\bm{a})^\dagger\Sigma_2^{(2)}(\bm{a})\Pi_2^{(2)}(z;\bm{a}) \, ,
\end{equation}
with the three-dimensional period vector
\begin{equation}
    \Pi_2^{(2)}(z,\bm{a})=
    \begin{pmatrix}
        \pcFq{3}{2}{a_3, \, 2-a_{245}, \, 1-a_5}{3-a_{1245}, \, 2- a_{25}}{z} \\
        z^{a_{25}-1}\pcFq{3}{2}{a_2, \, 1-a_{4}, \, -1+a_{235}}{2- a_{14}, \, a_{25}}{z} \\
        z^{a_{1245}-2}\pcFq{3}{2}{a_1, \, -1+a_{124}, \, -2+a_{12345}}{a_{14}, \, -1+a_{1245}}{z} 
    \end{pmatrix},
\end{equation}
and the intersection matrix
\begin{equation}
    \Sigma_2^{(2)}(\bm{a})=\begin{pmatrix}
        \frac{\mathfrak{s}(1-a_1)\mathfrak{s}(1-a_2)\mathfrak{s}(1-a_5)\mathfrak{s}(2-a_{245})}{\pi^2\mathfrak{s}(2-a_{25})\mathfrak{s}(3-a_{1245})} & 0 & 0 \\
        0 & -\frac{\mathfrak{s}(1-a_1)\mathfrak{s}(a_3)\mathfrak{s}(a_4)\mathfrak{s}(2-a_{235})}{\pi^2\mathfrak{s}(-1+a_{14})\mathfrak{s}(2-a_{25})} & 0 \\
        0 & 0 & -\frac{\mathfrak{s}(1-a_2)\mathfrak{s}(a_3)\mathfrak{s}(2-a_{124})\mathfrak{s}(3-a_{12345})}{\pi^2\mathfrak{s}(1-a_{14})\mathfrak{s}(3-a_{1245})} 
    \end{pmatrix} \, .
\end{equation}
Applying the recursion operator we find in four dimensions
\begin{equation}
    \phi_{4;\bm{a}}^{(2)}(z,\zb)=\frac{1}{(a_1-1)(a_2-1)(a_3-1)}\frac{1}{z-\zb}\Pi_4^{(2)}(z;\bm{a})^\dagger\Sigma_4^{(2)}(\bm{a})\Pi_4^{(2)}(z;\bm{a}) \, ,
\end{equation}
with the period vector
\begin{equation}
    \Pi_4^{(2)}(z;\bm{a})=
    \begin{pmatrix}
        \pcFq{3}{2}{-1+a_3, \, 3-a_{245}, \, 1-a_5}{5-a_{1245}, \, 3-a_{25}}{z} \\
        z\, \pcFq{3}{2}{a_3, \, 4-a_{245}, \, 2-a_5}{6-a_{1245}, \, 4-a_{25}}{z} \\
        z^{a_{25}-2}\,\pcFq{3}{2}{-1+a_2, \, 1-a_4, \, -3+a_{235}}{3-a_{14}, \, -1+a_{25}}{z} \\
        z^{a_{25}-1}\,\pcFq{3}{2}{a_2, \, 2-a_4, \, -2+a_{235}}{4-a_{14}, \, a_{25}}{z} \\
        z^{a_{1245}-4}\,\pcFq{3}{2}{-1+a_1, \, -3+a_{124}, \, -5+a_{12345}}{-1+a_{14}, \, -3+a_{1245}}{z} \\
        z^{a_{1245}-3}\,\pcFq{3}{2}{a_1, \, -2+a_{124}, \, -4+a_{12345}}{a_{14}, \, -2+a_{1245}}{z}
    \end{pmatrix}  \, ,
\end{equation}
and the intersection matrix given by
\begin{equation}
    \Sigma_4^{(2)}(\bm{a}) =
    \begin{pmatrix}
        0 & \sigma_1 & 0 & 0 & 0 & 0 \\
        -\sigma_1 & 0 & 0 & 0 & 0 & 0 \\
        0 & 0 & 0 & \sigma_2 & 0 & 0 \\
        0 & 0 & -\sigma_2 & 0 & 0 & 0 \\
        0 & 0 & 0 & 0 & 0 & \sigma_3 \\
        0 & 0 & 0 & 0 & -\sigma_3 & 0
    \end{pmatrix},
\end{equation}
with the non-trivial entries
\begin{align}
    \sigma_1&=\frac{(a_3-1)\fs(2-a_1)\fs(2-a_2)\fs(1-a_5)\fs(3-a_{245})}{\pi^2 \fs(3-a_{25})\fs(5-a_{1245})} \, , \nonumber \\
    \sigma_2&=-\frac{(a_2-1)\fs(2-a_1)\fs(-1+a_3)\fs(a_4)\fs(4-a_{235})}{\pi^2 \fs(-2+a_{14})\fs(3-a_{25})} \, , \\
    \sigma_3&=-\frac{(a_1-1)\fs(2-a_2)\fs(-1+a_3)\fs(4-a_{124})\fs(6-a_{12345})}{\pi^2 \fs(2-a_{14})\fs(5-a_{1245})} \, . \nonumber
\end{align}
As in the example of the box integral we could expand around the unit-propagator limit and find a double copy representation of the unit-propagator two-loop integral, which is given by a single-valued polylogarithm of weight 4. 
\subsection{Example: Window Integral}
\label{subsec:windowDoubleCopy}

\begin{figure}[t]
    \centering
    \includegraphics[]{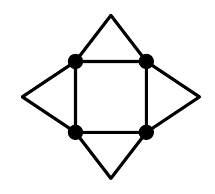}
    \caption{The four-loop window integral named after its similarity to a window in dual momentum space.}
    \label{fig:Window}
\end{figure}

As a final, less trivial, example let us study the unit-propagator power window integral, see \Figref{fig:Window}. Using the Basso--Dixon formula \cite{Basso:2017jwq} it can be expressed in terms of a determinant of ladder integrals
\begin{equation}
    \phi^{(2,2)}_{D=4,\gamma=1}=\frac{1}{12}\frac{1}{(z-\zb)^2}(f_1f_3-f_2^2)\,.
\end{equation}
Recall that the (rescaled) ladder functions were defined in eq.~\eqref{eq:rescaledLadder}. We showed above that the four-dimensional ladder integrals (for generic $\gamma$) inherit a double-copy form from the two-dimensional ones, which carries over into a (sum of terms in) double copy form for $\gamma=1$, as explicitly illustrated for the box integral above. These can of course again be embedded into a single-term double copy form involving a higher-dimensional period vector and intersection matrix. We will write this single-term double copy as
\begin{equation}
    f_k=\bar{\pi}_i^{(k)}\Sigma_{ij}^{(k)}\pi_j^{(k)} \,.
\end{equation}
Then the window integral (up to a prefactor) takes the form
\begin{align}
    f_1f_3-f_2^2=\bar{\pi}_i^{(1)}\bar{\pi}_j^{(3)}\Sigma_{ik}^{(1)}\Sigma_{jl}^{(3)}\pi_k^{(1)}\pi_l^{(3)}-\bar{\pi}_i^{(2)}\bar{\pi}_j^{(2)}\Sigma_{ik}^{(2)}\Sigma_{jl}^{(2)}\pi_k^{(2)}\pi_l^{(2)} \,.
\end{align}
This is clearly again a sum of factorized terms and can hence be written in an (even larger) double copy form
\begin{equation}
    \phi^{(2,2)}_{D=4,\gamma=1}(z,\zb)=\frac{1}{12}\frac{1}{(z-\zb)^2}\Pi_{2,2}(z)^\dagger\Sigma_{2,2}\Pi_{2,2}(z) \,.
\end{equation}
This argument clearly also applies to all other Basso--Dixon integrals, providing a new perspective on the well-known single-valuedness of this class of integrals. It would be interesting to explore in the future if this single-valuedness can also be shown for integrals involving position space loops in the absence of a determinant formula.
\section{Conclusions and Outlook}
\label{sec:Outlook}

In the present paper we have considered conformal Feynman integrals with a particular emphasis on four external points. We have shown that track-like integrals, composed of integration vertices arranged on a line, obey a dimensional recursion which, in particular, allows to lift their double copy structure from two to higher even dimensions. For integrals beyond four external points this statement only applies to a 2d subslice of the full kinematic space. Moreover, it was demonstrated that the operator which generates a loop recursion for four-point ladder integrals, induces Toda-like equations on the family of Basso--Dixon integrals.
\medskip

The above findings suggest a large number of interesting research directions. In particular, the present work shows close relations among the mathematical structures and techniques employed in the context of Feynman integrals, conformal field theory and integrability with large potential for furhter tightening these bonds. Let us list a selection of concrete follow-up questions.
\medskip

Firstly, the observation that the double copy form of Feynman integrals in two and higher dimensions is closely related, should be further explored beyond four external points. As indicated at the end of \Secref{subsec:confSchwingerParam}, for five points the recursion is a priori blind to one of the higher-dimensional conformal variables corresponding to the difference of five and four cross ratios in four and two spacetime dimensions, respectively. This suggests to expand the 4d result around the four-variable configuration that follows from the recursion from two dimensions, and to investigate the resulting mathematical structure in order to resolve the additional kinematic dependence systematically.
Note in this context that the two-loop ladder integral considered in \Secref{sec:2loopLadder} corresponds to a two-point coincidence limit of the $D$-dimensioanl 5-point conformal partial wave in the comb channel, which was considered in \cite{Rosenhaus:2018zqn} and can in turn be identified with a coincidence limit of the two-loop track (or double box) integral. Hence, as a next step one should extend the double copy form given below to the full 5-point kinematics following the above logic.
\medskip

Note that the fact that the dimensional recursion discussed in this paper only reaches a subspace of the full kinematic space does not need to be a drawback. In particular, in the context of defect or boundary conformal field theory, Feynman integrals evaluated on lower dimensional kinematics are of high interest. As a first step into this direction it will be interesting to use the recursion to reproduce the result for the double box integral in line-kinematics as given in \cite{Rodrigues:2024znq}. This can be considered as a higher-point generalization of the applications to polylogarithmic ladders presented in \Secref{subsec:SOVRep} above.
\medskip

On more general grounds, the relation between Feynman integrals and conformal blocks (or conformal partial waves) needs to be better understood. In particular, the expansion of fishnet Feynman integrals (alias correlation functions) into conformal blocks, which in 2d are well known to factorize into 1d conformal blocks \cite{Fortin:2020zxw}, should be related to the intersection pairing considered in \Secref{sec:doubleCopy}. This should e.g.\ relate the coefficients in expressions like \eqref{eq:shadowBlockDecomposition} for the box integral to the entries of the intersection matrix discussed in \Secref{subsec:boxDoubleCopy}.
\medskip

The holomorphic factorization of 2d integrals was of great use in \cite{Duhr:2022pch,Duhr:2023eld,Duhr:2024hjf} to compute multi-loop integrals based on their Yangian symmetry and Calabi--Yau geometry. The above indications on how this double copy structure lifts to higher dimensions should be further explored and exploited to combine them with the Yangian bootstrap for Feynman integrals in higher dimension \cite{Loebbert:2019vcj,Loebbert:2020glj,Corcoran:2020epz}.
\medskip

There is a well known relation between the double box integral in four spacetime dimensions and the six-dimensional hexagon~\cite{Paulos:2012nu} (see also \cite{Loebbert:2019vcj} for generalized propagator powers), which can be extended to other integrals as well, see e.g.~\cite{Spiering:2024sea}. Here both, dimension and loop order are shifted at the same time. In particular, the recursion generator lowers the loop order, thus mapping a more complicated integral to a simpler one. It would be interesting to also understand these relations within the present context.
In a similar direction in \cite{Corcoran:2021gda} it was shown that the second-order differential generators of the conformal Yangian algebra induce dimension shift relations on Basso--Dixon integrals which should be embedded into the above discussions. 
\medskip

The appearance of Toda equations and tau-functions for Basso--Dixon integrals discussed above is another example of the curious relation between quantum and classical integrable systems \cite{tauFunctionsBook}. These should be considered in the light of determinant formulas for various observables in superconformal quantum field theories, see e.g.\ \cite{Beccaria:2022ypy,Bajnok:2024epf}. It would be fascinating to identify the general pattern behind these findings and to turn them into a computational tool. In the present context it would be particularly useful to systematically understand the emergence of  Toda-equations starting from the spacetime representation of Feynman integrals, which would allow to rederive the Basso--Dixon formula as a solution to these equations. It would also be intriguing to identify further families of Feynman integrals with a determinant representation and to reveal a generic group theoretic origin.
\medskip

Finally, we believe that there is great potential for further unifying the concepts employed in the computation of Feynman integrals, conformal correlation functions and observables within integrable models. The four-point examples discussed in the present paper provide a natural ground for pinning down their common mathematical structures as well as for generalizations to higher numbers of external points.


\section*{Acknowledgements} 
We are grateful to
Claude Duhr,
Vasco Gon\c{c}alves,
Sylvain Lacroix,
Franziska Porkert and
Evgeny Sobko for discussions about different aspects of this paper. Moreover we would like to thank Claude Duhr for helpful comments on the manuscript.
\appendix
\section{Single-Valued Hypergeometric Functions}
\label{sec:svHGFs}
In this appendix we will explicitly define the single-valued generalized hypergeometric functions ${}_{p+1} \mathcal{F}^{\mathrm{sv}}_p(\bm{\alpha},\bm{\beta},z)$. The material here was worked out in \cite{svpFq1,svpFq2} and we are closely following \cite{Duhr:2023bku}.

We define the (rescaled) holomorphic generalized hypergeometric functions by
\begin{align}
    {}_{p+1}\mathcal{F}_p(\bm{\alpha},\bm{\beta},z)&=\left(\prod_{i=1}^p\frac{\Gamma(\alpha_i)\Gamma(\beta_i-\alpha_i)}{\Gamma(\beta_i)}\right) {}_{p+1}F_p(\bm{\alpha},\bm{\beta},z)  \\
    &=\prod_{i=1}^p\left( \int_0^1\dd t_i\, t_i^{\alpha_i-1}(1-t_i)^{\beta_i-\alpha_i-1} \right)(1-zt_1\dots t_p)^{-\alpha_0} \, ,
\end{align}
where we abbreviated $\bm{\alpha}=(\alpha_0,\alpha_1,\dots,\alpha_p),\, \bm{\beta}=(\beta_1,\dots,\beta_p)$.

The single-valued generalized hypergeometric functions can then be written as bilinears in the (anti-)holomorphic ones
\begin{equation}
    {}_{p+1} \mathcal{F}^{\mathrm{sv}}_p(\bm{\alpha},\bm{\beta},z)=\Pi_p(z;\bm{\alpha},\bm{\beta})^\dagger \Sigma_p(\bm{\alpha},\bm{\beta}) \Pi_p(z;\bm{\alpha},\bm{\beta})
\end{equation}
with the intersection matrix
\begin{align}
    \left(\Sigma_p(\bm{\alpha},\bm{\beta})\right)_{0,0}&=\prod_{i=1}^p\frac{\sin(\pi\alpha_i)\sin(\pi(\beta_i-\alpha_i))}{\pi\sin(\pi\beta_i)} \, ,\nonumber \\
    \left(\Sigma_p(\bm{\alpha},\bm{\beta})\right)_{i,i}&=-\frac{\sin(\pi\alpha_0)\sin(\pi(\beta_i-\alpha_0))}{\pi\sin(\pi\beta_i)}\prod_{j=1,j\neq i}^p\frac{\sin(\pi(\beta_i-\alpha_j))\sin(\pi(\beta_j-\alpha_i))}{\pi\sin(\pi(\beta_i-\beta_j))},\quad i>0 \, ,\nonumber\\
    \left(\Sigma_p(\bm{\alpha},\bm{\beta})\right)_{i,j}&=0,\quad i\neq j \, ,
\end{align}
and the period vector
\begin{align}
    \left(\Pi_p(z;\bm{\alpha},\bm{\beta})\right)_0&={}_{p+1}\mathcal{F}_p(\bm{\alpha},\bm{\beta},z) \, , \\
    \left(\Pi_p(z;\bm{\alpha},\bm{\beta})\right)_i &=z^{1-\beta_i}{}_{p+1}\mathcal{F}_p(\bm{\alpha}_i,\bm{\beta}_i,z),\quad i>0 \, ,
\end{align}
where we introduced the abbreviations
\begin{align}
    \alpha_i&=(1+\alpha_i-\beta_i,1+\alpha_1-\beta_i,\dots, \widehat{1+\alpha_i-\beta_i},\dots, 1+\alpha_p-\beta_i,1+\alpha_0-\beta_i)\, , \\
    \beta_i&=(1+\beta_1-\beta_i,\dots,\widehat{1+\beta_i-\beta_i},1+\beta_p-\beta_i,2-\beta_i) \, ,
\end{align}
for $i>0$. Here the hat over an entry means that it is omitted.

These functions are called single-valued since they are invariant under all monodromies. To make this explicit consider the period vector $\Pi(z)$ and analytically continue it along some closed contour $C$.
 Then the periods will in general pick up some monodromies, however this new period vector satisfies the same differential equations as the original one so they are related by a constant transformation, the monodromy matrix
\begin{equation}
    \Pi\rightarrow M_C\Pi \,.
\end{equation}
The special property of the single-valued hypergeometric functions is now that they are given by a combination of the periods which is invariant under all monodromies. This is achieved by the intersection form being invariant under all possible monodromy matrices $M_C$
\begin{equation}
    M_C^\dagger\Sigma M_C=\Sigma \,.
\end{equation}
In this sense the single-valued hypergeometric functions are invariant under monodromies and hence single-valued.

\bibliographystyle{nb}
\bibliography{FourPointInts}

\end{document}